\documentclass[prb,twocolumn,showpacs]{revtex4}
\usepackage{amssymb}

\usepackage{graphicx,amsmath}

\newcommand{\bea}{\begin{eqnarray}}
\newcommand{\eea}{\end{eqnarray}}
\newcommand{\ben}{\begin{equation}}
\newcommand{\een}{\end{equation}}
\newcommand{\benu}{\begin{enumerate}}
\newcommand{\enu}{\end{enumerate}}
\newcommand{\la}{\langle}
\newcommand{\ra}{\rangle}
\newcommand{\al}{\alpha}
\newcommand{\be}{\beta}
\newcommand{\ga}{\gamma}
\newcommand{\om}{\omega}
\newcommand{\Om}{\Omega}
\newcommand{\ep}{\epsilon}
\newcommand{\ta}{\tau}
\newcommand{\si}{\sigma}
\newcommand{\dl}{\delta}

\newcommand{\tht}{\theta}

\newcommand{\ord}{\mathcal{O}}
\newcommand{\mG}{\mathcal{G}}
\newcommand{\vol}{\mathcal{V}}

\newcommand{\ptl}{\partial}

\newcommand{\tOm}{\tilde{\Omega}}
\newcommand{\tS}{\tilde{S}}
\newcommand{\prf}{\mathcal{P}}
\newcommand{\eprm}{E^{\prime}}

\newcommand{\psid}{\psi^{\dagger}}

\newcommand{\bk}{{\bf k}}
\newcommand{\bq}{{\bf q}}
\newcommand{\bp}{{\bf p}}
\newcommand{\br}{{\bf r}}

\begin{document}

\title{Interaction Correction of Conductivity Near a Ferromagnetic
Quantum Critical Point}
\date{\today}
\author{I. Paul}

\affiliation{Materials Science Division, Argonne National Laboratory, Argonne,
IL 60439, USA}

\begin{abstract}
We calculate the temperature dependence of conductivity due to interaction correction
for a disordered itinerant electron system close to a ferromagnetic quantum critical point
which occurs due to a spin density wave instability.
In the quantum critical regime, the crossover between diffusive and
ballistic  transport occurs at a temperature
$T^{\ast }=1/[\tau \gamma (E_{F}\tau )^{2}]$, where $\gamma $ is the parameter
associated with the Landau damping of the spin fluctuations, $\tau $ is the
impurity scattering time, and $E_{F}$ is the Fermi energy. For a generic
choice of parameters, $T^{\ast }$ is few orders of magnitude
smaller than the usual crossover scale $1/\tau $. In the ballistic
quantum critical regime, the conductivity has a $T^{(d-1)/3}$
temperature dependence, where $d$ is the dimensionality of the system.
In the diffusive quantum critical regime we get $T^{1/4}$
dependence in three dimensions, and $\ln^2 T$ dependence in
two dimensions. Away from the quantum critical regime we recover the standard
results for a good metal.
\end{abstract}
\maketitle

\section{Introduction}
\label{sec:intro}

The effect of disorder and interaction on the temperature dependence of
conductivity of metals has been a topic of theoretical and experimental
investigations for over two decades.~\cite{altshuler,aleiner1,zna}
However, most of these
studies are on systems which are ``good metals" that behave as Fermi liquids (FLs),
for which the electron-electron interaction is short-ranged.
More recently, the observation of anomalous transport properties of metals which are
near putative quantum critical points (QCPs)~\cite{stewart}
has inspired theorists to examine the interplay
of disorder and interaction on transport properties of metals near
quantum criticality.~\cite{ipaul,kim,prl85,belitz2,rosch1}

In contrast with good metals, the electron-electron interaction near a QCP can be
long-ranged, which raises the possibility that in the latter case the
combined effect of disorder and interaction strongly influences the temperature
dependence of conductivity. From this perspective the study of charge transport near
a ferromagnetic QCP is particularly interesting.~\cite{ipaul}
Close to a
ferromagnetic QCP of the spin density wave variety, the spin fluctuations
are gapless (i.e., long ranged), but they do not break any lattice symmetry. As a result,
the contribution to resistivity due to the inelastic scattering of the carriers with the
spin fluctuations in a
clean system (where effects of impurity can be neglected) is zero, unless
Umklapp processes are taken into account in order to relax momentum.  On the
other hand, in a dirty system the ``interaction" correction to the residual resistivity is
expected to become important, especially at low enough temperature when the
lattice mediated inelastic scattering with spin fluctuations is frozen out.
The interaction correction is the result of quantum interference between semiclassical
electron paths where, along one path electrons are scattered elastically by impurities
and along the second path they are scattered by the self-consistent potential of
Friedel oscillations.~\cite{zna}
The study of the temperature dependence of conductivity due to
this subtle quantum interference process for a system close to a ferromagnetic QCP
is the topic of this paper.

From the point of view of experiments, the existence of a ferromagnetic QCP is
currently a topic of investigation. In most three dimensional compounds, such as
UGe$_2$~\cite{huxley} and ZrZn$_2$,~\cite{uhlarz}
the ferromagnetic transition from the paramagnetic state
becomes first order as the Curie temperature is lowered by the application of
pressure. In two dimensions the most promising candidate for exhibiting ferromagnetic
type of quantum critical behaviour is the bi-layer material Sr$_3$Ru$_2$O$_7$
which undergoes metamagnetic transition in the presence of an external magnetic
field.~\cite{grigera}
Until recently, it was believed that the metamagnetic transition in this
 material could be tuned to a quantum critical end point for which a spin fluctuation
 type of  theory was considered appropriate.~\cite{schofield}
 However, recent experiments on cleaner
 samples reveal that the approach to the quantum critical end point is pre-empted
 by a new phase transition, whose origin is itself a subject of investigation
 currently.~\cite{grigera2}
 On the other hand, the fact that this new phase transition in Sr$_3$Ru$_2$O$_7$
 is pushed to higher temperature for samples which are cleaner, provides empirical
 evidence that
 it may be possible to stabilize
 a continuous ferromagnetic transition at low temperature
 by the deliberate introduction of disorder. This point of view is further
 supported by a recent study of ZrZn$_2$ with Nb doping
 (which presumably introduces more disorder compared to a pressure tuning), where a lowering of
 Curie temperature has been reported keeping the transition continuous down to the
 lowest measured transition temperature.~\cite{sokolov}

 On the theoretical side, the effect of quantum interference on the temperature
 dependence of conductivity of a disordered metallic system is well understood
 for the case when
 the system is away from any QCP and when the electron-electron interaction is
 short-ranged.~\cite{altshuler,lee}  The effect is more dramatic in lower dimensions, where the
 temperature ($T$)
 dependent correction to the residual resistivity exhibit singular behaviour. In
 particular, in two dimensions the correction is logarithmic in $T$ in the diffusive
 regime when $T \tau \ll 1$,~\cite{altshuler}
 and linear in $T$ in the ballistic regime $T \tau \gg 1$,~\cite{zna}
 where $\tau$ is the elastic scattering lifetime of the electrons. In contrast, in
 three dimensions the temperature dependence is $\sqrt{T}$ in the diffusive
 regime,~\cite{altshuler}
 and $T^2 \log (T)$  in the ballistic regime where it is difficult to
 distinguish it from  $T^2$ terms that arise due to ordinary FL corrections.
 Quantum correction to conductivity has also been studied for models of gauge
 fields interacting with charged fermions.~\cite{mirlin,khveshchenko,galitski}
 Close to a QCP the interaction between the electrons is long-ranged
 (the same happens when the interaction between fermions is mediated by a
 gapless gauge boson)
which makes it difficult to formulate a controlled theory. Consequently, there are
 relatively fewer studies of transport properties of metals near quantum
 criticality.~\cite{ipaul,kim,prl85,belitz2,rosch1}
 For a
metamagnetic QCP in two dimensions it was shown earlier that the conductivity in
the diffusive regime has $\ln^2 T$ dependence,~\cite{ipaul,kim} in contrast
with  the usual $\ln T$ dependence of a good metal.

A controlled study of the interaction correction to conductivity for
a two dimensional electron system close to a ferromagnetic QCP
was performed in Ref.~\onlinecite{ipaul}. In this work
two new effects were identified which arise when the system is close to the QCP.
First, the
crossover between diffusive and ballistic regimes of transport near the QCP
occurs at a temperature $T^{\ast} = 1/[\tau \gamma (E_F \tau)^2]$, where
$\gamma$ is the parameter associated with the Landau damping of the spin fluctuations,
and $E_F$ is the
Fermi energy. For a generic choice of parameters, $T^{\ast}$ is much smaller than
the crossover scale $1/\tau$ which is expected in the case of a good metal.  Second,
in the ballistic quantum critical regime the temperature  dependence of
conductivity ($\sigma$)
has a new exponent, namely $\delta \sigma (T) \equiv \sigma (T) - \sigma (0)
\propto - T^{1/3}$.

In the current paper we extend the work of
Ref.~\onlinecite{ipaul}  to
study the interaction corrections for a three dimensional electron system near a
ferromagnetic QCP, and  we also provide some technical details  which are absent in
Ref.~\onlinecite{ipaul}. For the three dimensional case our main results are :
(i) in the quantum
critical regime the crossover between ballistic and diffusive transport occurs at a
temperature $T^{\ast} \ll 1/\tau$ (same as in two dimensions),
(ii) in the diffusive
quantum critical regime $\delta \sigma \propto - T^{1/4}$, and (iii) in the ballistic
quantum critical regime $\delta \sigma \propto - T^{2/3}$. Moving away from the
quantum critical regime we recover the usual results for a FL.
We note that in both two and three dimensions, and in all the crossover regimes
considered here, we find that the temperature dependence of conductivity due to interaction
correction has a metallic sign (i.e., $d \si/d T < 0$).

The organization of the rest of the paper is as follows. In section~\ref{sec:model}
we describe the model, and we explain the various technical steps that are
involved in the calculation of the conductivity. Some details of the calculations
are given in appendix~\ref{appen:a}. In section~\ref{sec:2d} we obtain the leading temperature
dependence of conductivity in the various crossover regimes for dimension $d=2$.
This section is an extended version of Ref.~\onlinecite{ipaul}.
In section~\ref{sec:3d} we calculate the leading temperature dependence of conductivity
in the various crossover regimes for dimension $d=3$. In section~\ref{sec:conclusion}
we conclude with a summary of our results.

\section{Model and Formalism}
\label{sec:model}
In the conventional method for studying quantum criticality in itinerant electron
systems the conduction electrons are formally integrated out, and a Landau-Ginzburg action
in terms of the order parameter fields is studied.~\cite{hmm}
Recently, the validity of integrating out low-energy electrons has been questioned,
and it has been argued that such a procedure generates singularities to all orders
in the collective spin interactions.~\cite{belitz1,abanov,belitz-rmp}
In the following we start with the phenomenological
spin-fermion model introduced in Ref.~\onlinecite{abanov}, which describes the low-energy properties
of electrons close to a ferromagnetic instability of the spin density wave type
in dimensions $d$ ($=2,3$), and add scattering of electrons due to static
impurities.  This is described by the action
\bea
\label{eq:action}
S &=&
T \sum_{\om_n} \int d^d r \psid_{\al} (\br, \om_n)
\left[ i \om_n + \frac{\nabla^2}{2m} + \mu \right] \psi_{\al} (\br, \om_n)
\nonumber \\
&+&
(E_0 T) \sum_{\Om_n \bq} U^{-1}(\bq, \Om_n) {\bf S} (\bq, \Om_n) \cdot
{\bf S} (- \bq, - \Om_n)
\nonumber \\
&+& \! \!
\left(\frac{\al E_0}{\nu_0} \right)^{1/2} \! \! \!
\int
\! \!
d^d r
\! \!
\int_0^{\be}
\! \! \! \!
d \ta \psid_{\al} (\br, \ta)
\psi_{\be} (\br, \ta) \left[ {\bf S} (\br, \ta) \cdot {\bf \si}_{\al \be} \right]
\nonumber \\
&+&
\int d^d r \int_0^{\be} d \ta \psid_{\al} (\br, \ta) V(\br)
\psi_{\al}(\br, \ta),
\eea
where summation over repeated indices is implied.
($\psid_{\al}$, $ \psi_{\al}$) are Grassman fields describing low-energy
electrons with spin $\al$ and mass $m$, ${\bf S} (\bq, \Om_n)$ is a bosonic field
describing collective spin fluctuations in the system, $E_0$ is an
associated energy scale, ${\bf \si}$ are Pauli matrices,
 $\nu_0$ is the density of states of non-interacting
electrons with spin at the Fermi level, $\mu$ is the chemical potential
and $\be$ is inverse temperature.
$\nu_0 = m/\pi$ for $d=2$, and
$\nu_0 = (p_F m)/\pi^2$ for $d=3$, where $p_F$ is the Fermi momentum.
The fields
${\bf S} (\bq, \Om_n)$ are obtained by formally integrating out electrons above a
certain energy cut-off, for example, below which the dispersion of
the electrons can be linearized. The disorder potential $V(\br)$ is assumed
to obey Gaussian distribution with $\la V(\br_1) V(\br_2) \ra
= \dl (\br_1 - \br_2)/ (2 \pi \nu_0 \ta)$. The dimensionless coupling constant
describing interaction between the electrons and the spin fluctuations is
taken to be $\al < 1$.

In the current model the dynamics of the spin fluctuations is overdamped because
their spectrum falls within the continuum of the
particle-hole excitations of the electrons (Landau damping).
This overdamping of the spin fluctuations, which is a consequence of the spin-fermion coupling,
can be thought of as the self-energy of the spin fluctuations which we have introduced at the
very beginning in the phenomenological model described by Eq.~(\ref{eq:action}).~\cite{abanov, chubukov1}
\begin{figure}[tbp]
\begin{center}
\includegraphics[width=8.8cm]{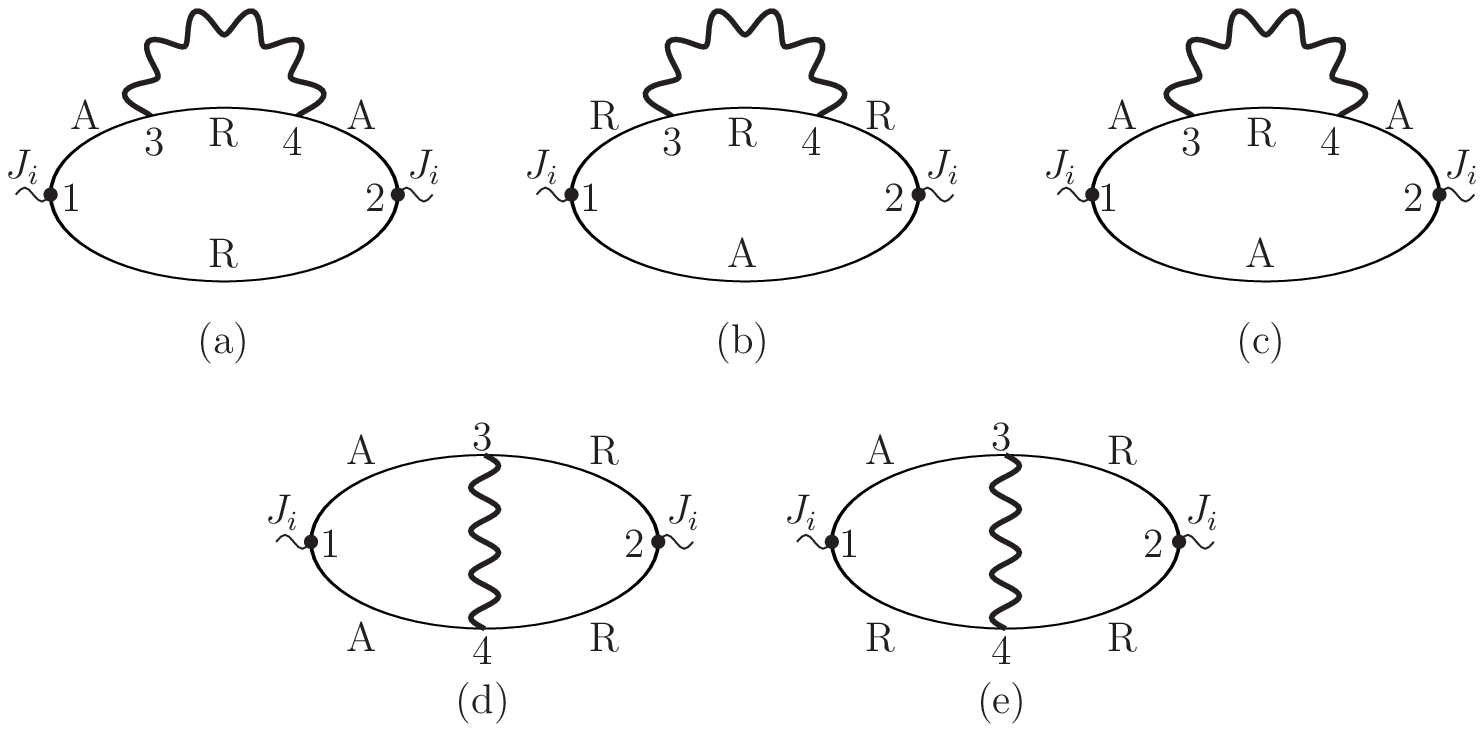}
\end{center}
\caption{Diagrams for interaction corrections to conductivity
at lowest order in interaction.~\cite{zna} The solid lines denote electron
Green's functions before disorder average. After disorder average
the same diagrams, with the solid lines now denoting disorder averaged
electron Green's functions, are
important in the ballistic regime. The wavy lines denote the propagator for
the spin fluctuation.}
\label{fig:cond1}
\end{figure}
 In this sense,
the current approach is analogous to a renormalized perturbation theory.
In the ballistic
regime the spin fluctuation is described by
\ben
\label{eq:Ub}
U(\bq, i \Om_n) \approx U_b(\bq, i \Om_n) =
\!
\left[  \dl + \! \left(  \frac{q}{p_F} \! \right)^2
\! \!
+ \frac{\ga \left| \Om_n \right|}
{v_F q} \! \right]^{-1},
\een
where $v_F = p_F/m$.
Here $\dl$ is the mass of the spin fluctuations which is related to the
magnetic correlation length $\xi$ by $\dl = (p_F \xi)^{-2}$, such that at the QCP
$\dl =0$. The dimensionless parameter $\ga$ is associated with the rate of
Landau damping which, in principle, is related to the coupling $\al$. For
example, $\ga$ should vanish when $\al$ is zero, while within random phase
approximation one gets $\ga = \al$.~\cite{chubukov1}
However, the precise relation between the
two parameters depends on microscopic details. In the following we consider
$\ga$ as an independent phenomenological parameter, and we find that with the
assumption
\ben
\label{eq:inequality}
\ga \gg \al
\een
it is possible to perform controlled calculation in
the entire $T$-$\dl$ plane. It is important to note that the form of the
Landau damping in the above Eq. is valid only in the quasi-static limit where
$v_F q \gg \Om$. In this limit the form is robust and is a universal feature
of the low-energy electrons.~\cite{agd}
In the opposite limit of $\Om \gg v_F q$ the damping term depends on
microscopic details, and the spin-fermion model as such loses universality.
In our calculation we find either the validity of the quasi-static limit,
thereby
justifying the universal form of the damping, or that the dynamics of the spin
fluctuations is unimportant to leading order. In this sense the results that
we derive in the following two sections are universal. In the diffusive regime
the damping term is modified because the particle-hole excitations generated
by the dissociation of the spin fluctuations have their own dynamics governed by the
diffusion pole.
In this regime we have
\ben
\label{eq:Ud}
U(\bq, i \Om_n) \approx U_d(\bq, i \Om_n) =
\!
\left[ \dl + \! \left(\frac{q}{p_F} \! \right)^2
\! \!
+ \frac{\ga \left| \Om_n \right|} {Dq^2} \! \right]^{-1},
\een
where $D = v_F^2 \ta/d$ is the diffusion constant.

For a system of electrons with short-ranged interaction,
i.e., when the system is well away from any phase instability,
it has been pointed out that the small $\bq$-expansion of the static spin
susceptibility starts with a non-analytic $\left| q \right|^{d-1}$ term for
dimension $d \leq 3$.~\cite{belitz1,infrared}
This non-analyticity is due to a $2 p_F$ singularity in the particle-hole
polarization function. At zero temperature, when the system is close to a
ferromagnetic instability, and in the absence of disorder, it has been shown
that the above non-analyticity changes into a $\left| q \right|^{(d+1)/2}$
term with a negative coefficient which favours either a first order transition
or a second order transition into a state with a finite ordering
wavevector.~\cite{rech}
On the other hand, in the presence of disorder the non-analyticity manifests
as a $\left| q \right|^{d-2}$ term.~\cite{disorder-belitz}
In the current study we neglect these non-analytic terms for the following reason.
We first note that, since the low-energy electrons are not integrated out in the
model given by Eq.~(\ref{eq:action}), in principle the above non-analytic terms are
also present in the current model. However, such terms are generated by higher order
spin-fermion coupling, and as such are sub-leading due to the condition given by
Eq.~(\ref{eq:inequality}). As we discuss later in this section, as well as in
appendices~\ref{appen:b} and~\ref{appen:c}, due to Eq.~(\ref{eq:inequality}) the
electron self-energy due to the spin-fermion coupling is sub-leading compared to their
elastic scattering rate. For the same reason, we find that the contributions to the
conductivity at second order in $\al$ are sub-leading as well. Consequently it is
reasonable to conjecture that, above a very low-temperature scale, the above
non-analytic terms can be ignored for the leading temperature dependence of the
conductivity. This is the justification for using the analytic $(q/p_F)^2$ terms
in Eqs.~(\ref{eq:Ub}) and~(\ref{eq:Ud}).

Close to the QCP there are two important temperature scales. (i) First,
$T^{\ast}$ which is defined as  the crossover
temperature between ballistic ($T \gg T^{\ast}$) and diffusive ($T \ll T^{\ast}$)
transport in the quantum critical regime.
The qualitative difference between these two regimes can be
understood as follows. Within the
time scale of an electron-electron interaction (mediated by the spin
fluctuations), if an electron undergoes typically a single
impurity scattering
then it corresponds to the ballistic limit. On the other hand, in the
diffusive regime an electron undergoes multiple impurity
scattering within that time scale. Using uncertainty relation, we estimate
the typical length travelled by an electron
during an electron-electron scattering event to be $1/q$, where $\bq$ is the
momentum transferred during interaction. This length scale becomes comparable
to the mean free path $v_F \ta$ at the crossover temperature $T^{\ast}$.
Close to the QCP [$\dl \ll (E_F \ta)^{-2}$, where $E_F$ is the Fermi energy]
the typical momentum transferred during
interaction is controlled by the pole in Eq.~(\ref{eq:Ub}), and is given by
$q_{B1} \sim p_F (\ga \Om/E_F)^{1/3}$. Scaling $\Om \sim T$, we get
$T^{\ast} \sim 1/[\ta (E_F \ta)^2 \ga]$. In the FL-regime far away from the
QCP ($\dl \gg \ga$), $q$ is determined by the typical momentum of a fermionic
excitation which is $q_F \sim \Om/v_F$, and the ballistic-diffusive crossover
scale is $1/\ta \gg T^{\ast}$. In the FL-regime near the QCP
[$(E_F \ta)^{-2} \ll \dl \ll \ga$], $ q \sim q_{B2} \sim (\ga \Om)/(v_F \dl)$
is still governed by the pole in Eq. (\ref{eq:Ub}), and the crossover scale
is $\dl/(\ga \ta)$. (ii) The second important temperature scale is
$T_1 = \ga^{1/2} E_F$, above which $q_F > q_{B1}$, and the effect of the
QCP on conductivity is wiped out by thermal fluctuations.

We get two possible situations
depending on the strength of the disorder characterized by
$1/(E_F \ta)$ relative to the Landau damping parameter $\ga$.
(a) For $\ga^{1/2} > 1/(E_F \ta)$, the low temperature
cutoff of the regime where $\dl \si \propto - T^{(d-1)/3}$ is $T^{\ast}$
and the high-$T$ cutoff is $T_1$. For $T < T^{\ast}$, we get
$\dl \si \propto \ln^2 (T)$ in d=2 (note our result has a metallic sign which
was missed in Ref.~\onlinecite{kim}), and $\dl \si \propto - T^{1/4}$ in $d=3$.
 (b) For $1/(E_F \ta) > \ga^{1/2}$, one get $T_1 < T^{\ast}$. In this
 situation the $T^{(d-1)/3}$ regime is lost and the effect of the QCP on conductivity
 is negligible. We note,
however, that the second situation is experimentally highly improbable for a good
metal for which $E_F \ta \sim 100$, while typically $\ga \sim 1$ (since it
is the ratio of the spin fluctuation velocity to the electron velocity). In
the rest of the paper we assume $\ga^{1/2} > 1/(E_F \ta)$ to be valid.

\begin{figure}[tbp]
\begin{center}
\includegraphics[width=6cm]{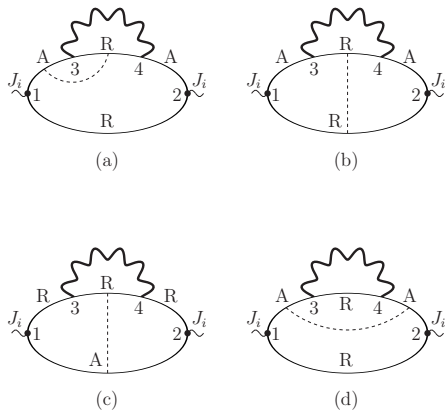}
\end{center}
\caption{Diagrams with explicit impurity scattering (dashed line)
for interaction corrections to conductivity
in the ballistic regime.}
\label{fig:cond2}
\end{figure}

In addition to scattering elastically with the impurity potential, the electrons also
scatter inelastically due to coupling with the spin fluctuations. At high enough
energies the inelastic scattering rate, given by the imaginary part of the
electron self-energy [${\rm Im} \Sigma (\om)$],
becomes larger than the
elastic scattering rate $1/\ta$, and in that case
the quantum interference effect is weak due to loss of phase coherence. In
$d=2$ one gets~\cite{rech}
${\rm Im} \Sigma^{(2d)} (\om - i \eta) \propto (\al E_F^{1/3}/\ga^{1/3}) \om^{2/3}$
from which, by comparing with $1/\ta$, we obtain a temperature scale
$T_x^{(2d)} = [\ga^{1/2}/(\al E_F \ta)^{3/2}]E_F$. For $\al > 1/(E_F \ta)$, one gets
$T_x^{(2d)} < T_1$, and in this case the high-$T$ cutoff of the regime with
$\dl \si \propto - T^{1/3}$ is determined by $T_x^{(2d)}$ instead of $T_1$.
In the opposite situation where $\al < 1/(E_F \ta)$, we get $T_x^{(2d)} > T_1$,
in which case the inelastic
scattering rate is significant only at high temperatures where quantum correction is anyway weak. In $d=3$ one gets
${\rm Im} \Sigma^{(3d)} (\om - i \eta) \propto \al |\omega|$, and a corresponding
$T_x^{(3d)} = 1/(\al \ta)$. In this case one can show (from $\ga^{1/2} > 1/(E_F \ta)$
and $\al < 1$) that $T_x^{(3d)} > T_1$ always.
For simplicity, in the following calculations we ignore the inelastic scattering rate
(i.e., electron self-energy and the associated scale $T_x$) entirely.

\begin{figure}[tbp]
\begin{center}
\includegraphics[width=10cm, trim= 10 0 -10 0]{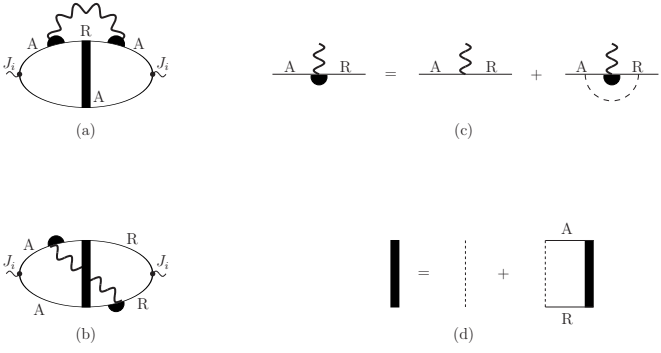}
\end{center}
\caption{Diagrams for interaction corrections to conductivity in the
diffusive regime.~\cite{altshuler}}
\label{fig:cond3}
\end{figure}

Next we discuss the technical details of the calculation of the interaction
corrections to conductivity. Within Kubo formalism the expression for
conductivity is given by
\[
\si_{ij} = \lim_{\Om \rightarrow 0} {\rm Im}
\left[ \frac{\Pi_{ij}(i \Om_n)}{i \Om_n} \right]_{i \Om_n \rightarrow
\Om + i \dl},
\]
where
\[
\Pi_{ij}(i \Om_n) =
\int_0^{\be}
d \ta \langle T_{\ta} \hat{J}_i (\ta)
\hat{J}_j (0) \rangle e^{i \Om_n \ta}
\]
is the current-current correlator.  The current operator is given by
\[
\hat{J}_i = \frac{i}{2m} \int d^d r \psid_{\al}(\br)
\left[ (\overrightarrow{\ptl}_r)_i - \overleftarrow{\ptl}_r)_i \right]
\psi_{\al}(\br),
\]
where ($i$, $j$) refer to spatial directions. The first step is to expand
the current-current correlator to the lowest order
in the interaction coupling $\al$.
We note that the vertex correction to the spin-fermion coupling, generated in
the next order in interaction, gives a sub-leading contribution to the interaction
correction of conductivity in the ballistic regime near the quantum critical point
(see appendix~\ref{appen:b}).
For
$T^{\ast} < T < T_1$, we find that the vertex correction is parametrically
small by $\al/\ga$ in $d=2$, and by $\al/\ga^{1/3}$ in $d=3$.
For $T > T_1$, the vertex contribution is small by
$(\al/\ga^{1/2})(T_1/T)^{1/3}$ in $d=2$, and by
$\al \ln (E_F/T)$ in $d=3$.
In the diffusive quantum critical regime the situation is the same
(see appendix~\ref{appen:c}).
We also note that in a low-energy effective model such as ours,
where the electron dispersion can be linearized, the Aslamazov-Larkin
contributions cancel out exactly due to particle-hole symmetry.~\cite{kamenev}
 The second step is to
perform analytic continuation in order to get the retarded
current-current correlator.
In this step the electron Green's functions that enter the
expression for the current-current correlator get continued into appropriate
combinations of retarded and advanced Green's functions. The expression for
the interaction correction to longitudinal conductivity is given by~\cite{zna}
\bea
\label{eq:cond0}
\dl \si_{ii}
\! \!
&=&
\! \!
- {\rm Im} \! \!
\int \! \! d^d r_1 \cdots d^d r_4 \! \left(\! \frac{3 \al}{\nu_0} \!\right)
\! \!
\int_{-\infty}^{\infty} \frac{d \Om}{4 \pi^2} \!
\left[ \frac{\ptl}{\ptl \Om} \!
\left( \! \Om \coth \frac{\Om}{2T} \! \right) \! \right]
\nonumber \\
&\times&
U^A_{34}(\Om) \left[ J_{1i} G^A_{13}(\om) G^R_{34}(\om -\Om) G^A_{42}(\om)
J_{2i} G^R_{21}(\om) \right.
\nonumber \\
&+&
J_{1i} G^R_{13}(\om) G^R_{34}(\om -\Om) G^R_{42}(\om) J_{2i} G^A_{21}(\om)
\nonumber \\
&-&
J_{1i} G^A_{13}(\om) G^R_{34}(\om -\Om) G^A_{42}(\om) J_{2i} G^A_{21}(\om)
\nonumber \\
&-&
J_{1i} G^A_{41}(\om) G^A_{13}(\om) G^R_{32}(\om -\Om) G^R_{24}(\om-\Om)J_{2i}
\nonumber \\
&+&
\left.
2 J_{1i} G^R_{41}(\om) G^A_{13}(\om) G^R_{32}(\om -\Om)
G^R_{24}(\om-\Om)J_{2i} \right],
\eea
where $G^{R,A}(\om)$ is the Green's function for non-interacting
electrons in the presence of a random potential (i.e., before disorder
average), and $J_i =(i/(2m))[\overrightarrow{\ptl}_i -
\overleftarrow{\ptl}_i ]$.
The corresponding diagrams are shown in Fig.~(\ref{fig:cond1}).
The third step is
to perform the disorder average which restores translation invariance. The
disorder averaged electron Green's function is given by
\ben
\label{eq:G-fn}
\langle G^{R,A} (\bk, \om) \rangle = \left(
\om - \ep_{\bk} \pm \frac{i}{2 \ta} \right)^{-1}.
\een
Here $\ep_{\bk}$ is the linearized electron dispersion as measured from the
Fermi energy. The diagrams which contribute in the ballistic regime are shown
in Figs.~(\ref{fig:cond1}, \ref{fig:cond2}), while those that are
important in the
diffusive regime are shown in Fig.~(\ref{fig:cond3}).
In these diagrams the electron propagator is denoted by a solid line, the
propagator for the spin fluctuations by a wavy line, and an explicit
impurity scattering (in contrast with the implicit ones which give elastic
scattering lifetime to the electrons in Eq.~(\ref{eq:G-fn})), which gives a
factor of $1/(2\pi \nu_0 \tau)$, by a dashed line.

The interaction correction to conductivity in the triplet channel
can be written as~\cite{zna}
\bea
\label{eq:cond1}
\dl \si_T
&=&
 - \left( \frac{6 \pi e^2 v_F^2 \ta \al}{d} \right)
\int_{-\infty}^{\infty}
\frac{d \Om}{4 \pi^2} \left[ \frac{\ptl}{\ptl \Om} \left( \Om \coth
\frac{\Om}{2T} \right) \right]
\nonumber \\
&\times&
{\rm Im} \int \frac{d^d q}{(2 \pi)^d} U^A(\bq, \Om) B^{(2d,3d)} (\bq, \Om),
\eea
where $U^A(\bq, \Om)$ is the advanced bosonic propagator
given by Eqs.~(\ref{eq:Ub}) and (\ref{eq:Ud}),
and $B^{(2d,3d)}(\bq, \Om)$ is the fermionic part in the diagrams shown in
Figs.~(\ref{fig:cond1}, \ref{fig:cond2}, \ref{fig:cond3}) in dimensions
two and three respectively. In the ballistic regime
$v_Fq \gg 1/\ta$, and therefore diagrams
with more than one explicit impurity scattering are sub-leading.
The limiting form of
$B$ in this regime is given by the leading term in $\ta$ from the sum of the
diagrams given by Figs. (\ref{fig:cond1}, \ref{fig:cond2}). The details of
this evaluation is given in Appendix \ref{appen:a}. This approximation is equivalent
to an expansion in $(T^{\ast}/T)^{1/3}$ near the QCP, and in $1/(T \ta)$ for
$\dl \gg \ga$ (i.e., in the FL regime). In this regime $B \approx B_b$, and in two dimensions we get
\ben
\label{eq:Bb2}
B_b^{(2d)}(\bq, \Om) = \frac{2}{(v_F q)^2} \left(1- \frac{i \Om}{S_0}
\right)^2
+ \frac{2}{S_0^2} \left( 1 - \frac{i \Om}{S_0} \right),
\een
where
\ben
\label{eq:S0}
S_0 = \left[ (v_F q)^2 - \Om^2 + i \eta {\rm Sgn}(\Om) \right]^{1/2},
\een
while in three dimensions we get
\ben
\label{eq:Bb3}
B_b^{(3d)} = \frac{2}{(v_F q)^2} \left(1 - \frac{i \Om}{S_1} \right)^2
+ \frac{2}{S_1^2} - \frac{2}{S_0^2},
\een
where
\ben
\label{eq:S1}
\frac{1}{S_1} = \frac{1}{2i v_F q} \ln \left[ \frac{\Om + v_F q - i \eta}
{\Om - v_Fq - i \eta} \right].
\een
In the diffusive regime $v_F q \ll 1/\ta$, and multiple impurity scattering
needs to be taken into account.~\cite{altshuler}
The leading contribution is given by the
diagrams in Fig.~(\ref{fig:cond3}), whose evaluation is discussed
in Appendix \ref{appen:a}. In this
limit the kinematics of the electrons is governed by the diffusion pole, and
we get $B \approx B_d$, where
\ben
\label{eq:Bd}
B_d^{(2d,3d)} (\bq, \Om) = \frac{4}{d^2}
\frac{\ta (v_F q)^2}{(i \Om + Dq^2)^3},
\een
in dimensions $d$.

\section{Results in two dimensions}
\label{sec:2d}

From Eq.~(\ref{eq:cond1}) the correction to conductivity in the triplet channel is
\bea
\dl \si_T
&=&
 - ( 3 \pi e^2 v_F^2 \ta \al)
\int_{-\infty}^{\infty}
\frac{d \Om}{4 \pi^2} \left[ \frac{\ptl}{\ptl \Om} \left( \Om \coth
\frac{\Om}{2T} \right) \right]
\nonumber \\
&\times&
{\rm Im} \int \frac{d^2 q}{(2 \pi)^2} U^A(\bq, \Om) B^{(2d)} (\bq, \Om),
\nonumber
\eea
where $U^A(\bq, \Om)$ is given by Eqs.~(\ref{eq:Ub}) and (\ref{eq:Ud}) in the ballistic
and diffusive regimes respectively, and $B^{(2d)}$ by Eqs.~(\ref{eq:Bb2})
and (\ref{eq:Bd}) respectively.

\subsection{Ballistic Regime}

\begin{figure}[tbp]
\begin{center}
\includegraphics[width=7.8cm]{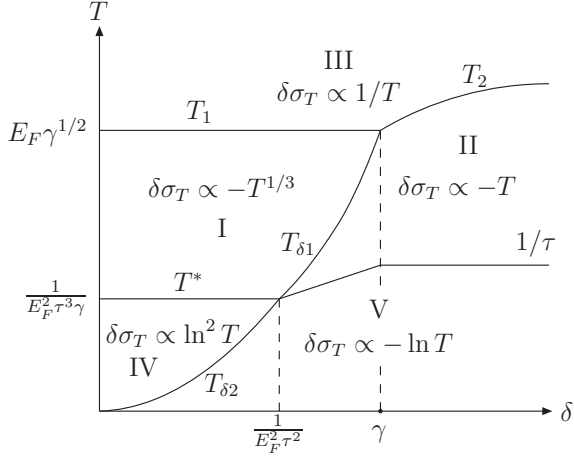}
\end{center}
\caption{Various crossover regimes for the leading temperature dependence of
the triplet channel contribution to conductivity in $d=2$.
$T_{\protect\delta 1}=(\protect\delta ^{3/2}/\protect\gamma
)E_{F}$, $T_{\protect\delta 2}=(\protect\delta ^{2}\protect\tau /\protect%
\gamma )E_{F}^{2}$, $T_{2}=E_{F}\protect\delta ^{1/2}$. Notice that $\protect%
\gamma ^{1/2}\gg 1/(E_{F}\protect\tau )$.}
\label{fig:scales1}
\end{figure}

The ballistic regime is defined by $T \gg T^{\ast}$ for $\dl \ll (E_F \tau)^{-2}$, by
$T \gg \dl/(\ga \tau)$ for $(E_F \tau)^{-2} \ll \dl \ll \ga$, and by $T \gg 1/\tau$ for
$\dl \gg \ga$. In this limit there are three crossover
regimes (regions I-III in Fig.~(\ref{fig:scales1})).

\emph{Regime I}. In order to calculate the leading behaviour in this regime
one can set $\dl =0$
in Eq.~(\ref{eq:Ub}), which gives the momentum scale $q_{B1} \sim p_F (\ga \Om/E_F)^{1/3}$.
This is the typical momentum transferred by the spin fluctuations to the electrons during
elastic scattering at temperature $T \sim \Om$. The high-$T$ cut-off of this regime is
$T_1 = \ga^{1/2} E_F$, below which $v_F q_{B1} > \Om$ (in other words
$q_{B1} > q_F \sim \Om/v_F$, where
$q_F$ is the typical momentum of the fermionic excitations). As a result, the
frequency dependence
of $B^{(2d)} \approx B_b^{(2d)}$ in Eq.~(\ref{eq:Bb2}) can be ignored, giving
$B_b^{(2d)} \approx 4/(v_F q)^2$.
With these approximations we get
\[
{\rm Im} \int \frac{d^2 q}{(2 \pi)^2} U^A_b B_b^{(2d)}
=
- \left(\frac{2 p_F^{2/3}}{3 v_F^{4/3} \ga^{2/3}} \right)
\frac{1}{\left| \Om \right|^{2/3}}{\rm Sgn} (\Om).
\]
The last frequency integral is ultraviolet divergent for  which we introduce a
cut-off at $p_F v_F \sim E_F$. We get,
\ben
\label{eq:2d-I}
\dl \si_T = - \frac{e^2 \tau \al}{\pi \ga^{2/3}} \mathcal{C}_1 (p_F v_F)^{2/3} T^{1/3},
\een
where
\[
\mathcal{C}_1 \equiv - \int_0^{\infty} \frac{\ptl}{\ptl t} \left( \frac{2t}{e^t -1} \right)
\frac{1}{t^{2/3}} \approx 3.47.
\]
In Eq.~(\ref{eq:2d-I}), and in subsequent evaluations,
we ignore a temperature independent contribution which
renormalizes the residual conductivity. At finite $\dl$ the regime ends when
$\dl \sim (q_{B1}/p_F)^2$, which gives the crossover temperature scale
$T_{\dl 1} = (\dl^{3/2}/\ga) E_F$. For $T < T_{\dl 1}$ the effect of finite $\dl$ cannot
be neglected.

The result in Eq.~(\ref{eq:2d-I}) can also be simply understood from the following scaling
argument. The correction to the transport scattering rate can be estimated as
$\Delta[1/\tau]^{(2d)} \sim (1/\tau) {\rm Im} \Sigma^{(2d)}(\om) \Delta t$, where
$\Sigma^{(2d)}(\om) \propto \om^{2/3}$ is the self-energy of the electrons due to interaction
with the spin fluctuations, and $\Delta t$ is the average time scale of the interaction. In
this estimate ${\rm Im} \Sigma$ is the quasiparticle scattering rate, which is renormalized by
the factor $(\Delta t)/\tau$ in order to obtain a transport rate. Using the uncertainty principle
we estimate $\Delta t \sim 1/(v_F q)$, where ${\bf q}$ is the typical momentum transferred during
scattering between the electrons and the spin fluctuations. Near the QCP, $q \sim \om^{1/3}$, and
scaling $\om \sim T$, we obtain the $T$-dependence in Eq.~(\ref{eq:2d-I}).

\emph{Regime II}. In this regime we identify two situations. (i) First, for
$(E_F \tau)^{-2} \ll \dl \ll \ga$, the approximate form of $U_b^A$ is given by dropping
the $(q/p_F)^2$ term in Eq.~(\ref{eq:Ub}), giving $U_b^A \approx [\dl + i \ga \Om/(v_F q)]^{-1}$.
The typical scale of momentum transferred during scattering is given by
$q_{B2} \sim (\ga \Om)/(v_F \dl)$. Since $q_{B2} > q_F$ in this sub-regime, the $\Om$-
dependence of $B_b^{(2d)}$ can be ignored giving $B_b^{(2d)} \approx 4/(v_F q)^2$.
We get
\[
{\rm Im} \int \frac{d^2 q}{(2 \pi)^2} U_b^A B_b^{(2d)}
=
- \frac{1}{v_F^2 \dl} {\rm Sgn} (\Om).
\]
(ii) Second, for $\dl \gg \ga$, the typical momentum scale is given by $q_F \sim \Om/v_F$.
In this sub-regime the Landau damping term in $U_b^A$ is order $\ga \ll \dl$, and so it can
be ignored giving $U_b^A \approx 1/\dl$ (the situation is non quasi-static since
$v_F q \sim \Om$, however the Landau damping term, whose universal form in Eq.~(\ref{eq:Ub})
is correct only in the quasi-static limit, becomes unimportant for leading behaviour). We
retain the full $\Om$-dependence of $B_b^{(2d)}$ given by Eq.~(\ref{eq:Bb2}), and we get
\[
\frac{1}{\dl} {\rm Im} \int \frac{d^2 q}{(2 \pi)^2} B_b^{(2d)}(\bq, \Om)
=
- \frac{1}{v_F^2 \dl} {\rm Sgn} (\Om),
\]
i.e., the same result as in sub-regime (i). For both the sub-regimes the triplet
channel contribution to conductivity is
\ben
\dl \si_T = - \left(\frac{3 e^2 \tau \al}{\pi \dl}\right) T,
\een
which is the result obtained in Ref.~\onlinecite{zna}.
The high-$T$ cut-off of sub-regime (ii) is given by $T_2 = \dl^{1/2} E_F$, above
which the $(q/p_F)^2$ term in $U_b^A$ cannot be neglected since
$(q/p_F)^2 \sim (\Om/E_F)^2 \gg \dl$.

\emph{Regime III}. This is the high temperature regime of the theory where the
typical momentum scale is $q_F \sim \Om/v_F$. For $\dl \ll \ga$, the damping
term in the spin fluctuation propagator can be neglected since
$(q/p_F)^2 \sim (\Om/E_F)^2 \gg \ga$. For $\dl \gg \ga$, the mass of the spin
fluctuations can be neglected since $(q/p_F)^2 \gg \dl$. Thus, in this regime
we get $U_b^A \approx (p_F/q)^2$. Since
\[
{\rm Im} \int \frac{d^2 q}{ (2 \pi)^2} \frac{p_F^2}{q^2} B_b^{(2d)}(\bq, \Om)
= 0,
\]
the leading order contribution to $\si_T$ cancel out, and $\dl \si_T \propto 1/T$
from sub-leading terms. In this regime the temperature dependence of conductivity
is dominated by the contribution from the singlet channel
$\dl \si \approx \dl \si_S \propto T$ (or by inelastic processes, if the
electron system is on a lattice).

\subsection{Diffusive Regime}

\emph{Regime IV}. Setting $\dl = 0$ we get
$U_d^A \approx [(q/p_F)^2 + i \ga \Om/(Dq^2)]^{-1}$,
which gives a possible momentum scale
$q_{D1} \sim p_F (\ga \Om/\eprm)^{1/4}$ where
$\eprm = Dp_F^2$. From the fermionic part $B_d^{(2d)}$
given by Eq.~(\ref{eq:Bd}), we get
a second momentum scale $q_{D2} \sim p_F(\Om/\eprm)^{1/2}$.
It can be shown that in this
regime $q_{D1} > q_{D2}$, which implies that the $\Om$-dependence
of $B_d^{(2d)}$ can be
ignored for the leading result, giving $B_d^{(2d)} \approx 2/(Dq^2)^2$.
With this approximation the resulting momentum integral is infrared divergent,
which is cut-off by the ignored momentum scale $q_{D2}$. Using
\bea
&{\rm Im}&  \int_{q_{D2}}^{\infty} \frac{d^2 q}{(2 \pi)^2}
\left[ \frac{q^2}{p_F^2} + \frac{i \ga \Om}{Dq^2} \right]^{-1}
 \frac{2}{(Dq^2)^2}
 \nonumber \\
 &=&
- \frac{1}{4 \pi \ga D \Om} \ln \left( \frac{\ga D p_F^2}{\left| \Om \right|}
\right),
\nonumber
\eea
we get
\ben
\dl \si_T = \left( \frac{3 e^2 \al}{8 \pi^2 \ga} \right) \ln ^2
\left( \frac{\ga D p_F^2}{T} \right).
\een
For finite $\dl$ this regimes exists for $ T > T_{\dl 2} = (\dl^2 D p_F^2)/\ga$. Below
$T_{\dl 2}$ the effect of finite $\dl$ cannot be neglected. This regime has been
 discussed previously in the context of $d=2$ metamagnetic QCP,~\cite{kim}
and also in the context of fermionic
gauge field models.~\cite{mirlin,khveshchenko}
Note that our result gives a metallic sign to the $T$-dependence of
conductivity.

\emph{Regime V}.
In this regime the $(q/p_F)^2$ term in $U_d^A$ can be dropped, giving
$U_d^A \approx [\dl + i \ga \Om/(Dq^2)]^{-1}$. This provides a new bosonic
momentum scale $q_{D3} \sim p_F [\ga \Om/(\eprm \dl)]^{1/2}$. As in Regime II,
two sub-regimes can be identified. (i) For $\dl \ll \ga$ we get $q_{D3} > q_{D2}$,
which implies that the $\Om$-dependence of $B_d^{(2d)}$ can be dropped, giving
$B_d^{(2d)} \approx 2/(Dq^2)^2$. Using $q_{D2}$ as an infrared cut-off
to the resulting momentum integral we get to leading order,
\[
{\rm Im} \int_{q_{D2}}^{\infty} \frac{d^2 q}{(2 \pi)^2}
\left[ \dl + \frac{i \ga \Om}{Dq^2} \right]^{-1}
\frac{2}{(Dq^2)^2}
= - \frac{1}{2 \pi D \ga \Om} \ln \left( \frac{\ga}{\dl} \right).
\]
(ii) For $\dl \gg \ga$ we get $q_{D2} > q_{D3}$. This implies that $U_d^A \approx 1/\dl$,
while the full $\Om$-dependence of $B_d^{(2d)}$ has to be retained. We get
\[
\frac{1}{\dl} {\rm Im} \int \frac{d^2 q}{(2 \pi)^2} B_d^{(2d)}(\bq, \Om)
= - \frac{1}{4 \pi \dl D \Om}.
\]
Combining the results of the two sub-regimes we get,
\ben
\dl \si_T = \frac{3 e^2 \al \mathcal{C}_2}{2 \pi^2 \ga} \ln
\left( \frac{E_F}{T} \right),
\een
where $\mathcal{C}_2 = \ln (\ga/\dl)$ for $\dl \ll \ga$,
and $\mathcal{C}_2 = \ga/(2 \dl)$ for $\dl \gg \ga$.
This is the famous Altshuler-Aronov correction to the conductivity
for the triplet channel in the diffusive regime of good metals.~\cite{altshuler}

\section{Results in Three Dimensions}
\label{sec:3d}

\begin{figure}[tbp]
\begin{center}
\includegraphics[width=7.5cm]{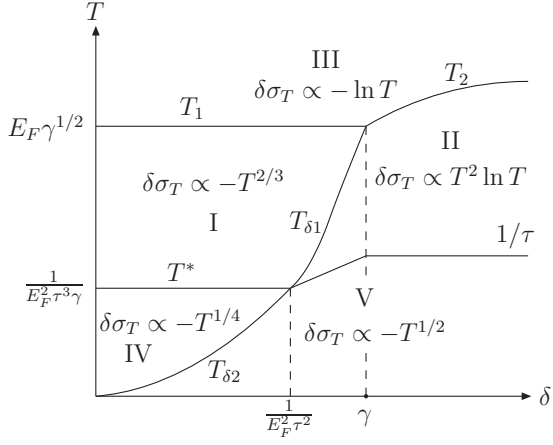}
\end{center}
\caption{Different crossover regimes for the temperature dependence of
the triplet channel contribution to conductivity in $d=3$.
$T_{\protect\delta 1}=(\protect\delta ^{3/2}/\protect\gamma
)E_{F}$, $T_{\protect\delta 2}=(\protect\delta ^{2}\protect\tau /\protect%
\gamma )E_{F}^{2}$, $T_{2}=E_{F}\protect\delta ^{1/2}$. Notice that $\protect%
\gamma ^{1/2}\gg 1/(E_{F}\protect\tau )$.}
\label{fig:scales2}
\end{figure}

From Eq.~(\ref{eq:cond1}) we get,
\bea
\label{eq:cond}
\dl \si_T
&=&
 - ( 2  \pi e^2 v_F^2 \ta \al)
\int_{-\infty}^{\infty}
\frac{d \Om}{4 \pi^2} \left[ \frac{\ptl}{\ptl \Om} \left( \Om \coth
\frac{\Om}{2T} \right) \right]
\nonumber \\
&\times&
{\rm Im} \int \frac{d^3 q}{(2 \pi)^3} U^A(\bq, \Om) B^{(3d)} (\bq, \Om),
\nonumber
\eea
where $B^{(3d)}(\bq, \Om)$ is given by Eqs.~(\ref{eq:Bb3}) and (\ref{eq:Bd}) in the ballistic
and diffusive regimes respectively. As in $d=2$, there are five
different crossover regimes (I---V in Fig.~(\ref{fig:scales2})) for the leading temperature
dependence of the correction to conductivity.
The scale of the typical momentum transferred by the spin fluctuations
to the conduction electrons during elastic scattering in each of these regimes
remain the same as in $d=2$. Consequently, the crossover lines delineating the five
regimes also remain the same as in $d=2$.

\subsection{Ballistic Regime}

\emph{Regime I}. Setting $\dl =0$ we get $U_b^A \approx [(q/p_F)^2 + (i \ga \Om)/(v_F q)]^{-1} $,
and the typical momentum scale is $q_{B1} \sim p_F (\ga \Om/E_F)^{1/3}$, so that
the $\Om$-dependence
of $B_b^{(3d)}$ can be dropped giving $B_b^{(3d)} \approx \pi^2/(2 v_F^2 q^2)$.
For the momentum integral we get
\bea
&{\rm Im}&  \int \frac{d^3 q}{(2 \pi)^3}
\left[ \frac{q^2}{p_F^2} + \frac{i \ga \Om}{v_F q} \right]^{-1}
\frac{\pi^2}{2 v_F^2 q^2}
\nonumber \\
&=&
-  \left( \frac{\pi}{12 \sqrt{3}} \right)
\frac{p_F^{4/3}{\rm Sgn} (\Om)}{v_F^{5/3} \ga^{1/3} \left| \Om \right|^{1/3}} .
\nonumber
\eea
The leading temperature dependence of the correction to conductivity is given by
\ben
\label{eq:3d-I}
\dl \si_T = - \frac{\mathcal{C}_3}{12 \sqrt{3}} \left( \frac{e^2 \tau \al p_F^{4/3} v_F^{1/3}}{\ga^{1/3}}
\right) T^{2/3},
\een
where
\[
\mathcal{C}_3 \equiv - \int_0^{\infty} dt \frac{\ptl}{\ptl t} \left( \frac{2t}{e^t -1} \right) \frac{1}{t^{1/3}}
\approx 2.21.
\]

As in the case of $d=2$, the result in Eq.~(\ref{eq:3d-I}) can be estimated from
$\Delta[1/\tau]^{(3d)} \sim (1/\tau) {\rm Im} \Sigma^{(3d)}(\om) \Delta t$.
Since ${\rm Im} \Sigma^{(3d)}(\om) \propto \om$, and $\Delta t \sim 1/(v_F q)$,
with $q \sim \om^{1/3}$ and $\om \sim T$, we get the exponent $2/3$ of the
temperature dependence in Eq.~(\ref{eq:3d-I}).

\emph{Regime II}.  As in $d=2$, two situations can be identified in this regime.
(i) For $(E_F \tau)^{-2} \ll \dl \ll \ga$, $U_b^A \approx [\dl + (i \ga \Om)/(v_F q)]^{-1}$, which
gives the momentum scale $q_{B2} \sim  (\ga \Om)/(v_F \dl)$. The $\Om$-dependence of
$B_b^{(3d)}$ can be dropped, and from the momentum integral
(which is ultraviolet divergent, and is cut-off at $p_F$) we get
\[
{\rm Im} \! \int  \! \frac{d^3 q}{(2 \pi)^3} \!
\left[ \dl + \frac{i \ga \Om}{v_F q} \right]^{-1}
\frac{\pi^2}{2 v_F^2 q^2}
= - \left( \frac{\ga \Om}{4 v_F^3 \dl^2} \right)  \ln \left(
\frac{\dl E_F}{\ga \left| \Om \right|} \right).
\]
(ii) For $\dl \gg \ga$ we have $U_b^A \approx 1/\dl$, and the typical
momentum scale is $q_F \sim \Om/v_F$. Keeping the full $\Om$-dependence of
$B_b^{(3d)}$ we get,
\[
\frac{1}{\dl} {\rm Im} \int \frac{d^3 q}{(2 \pi)^3} B_b^{(3d)}
= - \left( \frac{2 \Om}{\pi v_F^3 \dl} \right) \ln \left(
\frac{E_F}{ \left| \Om \right|} \right).
\]
After the $\Om$-integral we get
\ben
\dl \si_T = - \left( \frac{2 e^2 \tau \al \mathcal{C}_4}{3 v_F \dl} \right) T^2
\ln \left( \frac{\mathcal{C}_5 E_F}{T} \right),
\een
where $\mathcal{C}_4 = (\pi \ga)/((8 \dl)$ and $\mathcal{C}_5 = \dl/\ga$ for
$(E_F \tau)^{-2} \ll \dl \ll \ga$, and $\mathcal{C}_4 = \mathcal{C}_5 =1$ for
$\dl \gg \ga$.

\emph{Regime III}. In the high-$T$ regime of the theory $U_b^A \approx p_F^2/q^2$, and
the typical momentum scale is $q_F \sim \Om/v_F$. Keeping the full $\Om$-dependence of
$B_b^{(3d)}$ we get
\[
 {\rm Im} \int \frac{d^3 q}{(2 \pi)^3} \frac{p_F^2}{q^2}  B_b^{(3d)}
 = - \left( \frac{\mathcal{C}_6 p_F^2}{2 \pi v_F} \right) \frac{1}{\Om},
 \]
 where $\mathcal{C}_6 \equiv (4/3)\ln 2 - 1/3 \approx 0.60$.
 After the frequency integral we get,
 \ben
 \dl \si_T = \left( \frac{\mathcal{C}_6 e^2 \tau \al p_F^2 v_F}{2 \pi^2} \right) \ln \left(
 \frac{E_F}{T} \right).
 \een

 \subsection{Diffusive Regime}

 \emph{Regime IV}.  Setting $\dl = 0$ we get $U_d^A \approx [(q/p_F)^2 + i \ga \Om/(Dq^2)]^{-1}$,
 which gives the typical momentum scale $q_{D1} \sim p_F (\ga \Om/\eprm)^{1/4}$, where
 $\eprm = Dp_F^2$. Ignoring the $\Om$-dependence of $B_d^{(3d)}$ we have
 $B_d^{(3d)} \approx 4/[3 (Dq^2)^2]$. From the momentum integral we get
 \bea
\label{eq:t-1over4}
 &{\rm Im}& \int \frac{d^3 q}{(2 \pi)^3}
 \left[ \frac{q^2}{p_F^2} + \frac{i \ga \Om}{Dq^2} \right]^{-1}
 \frac{4}{3 (Dq^2)^2}
   \nonumber \\
 &=&
 - \frac{p_F^{1/2} {\rm Sgn}(\Om)}{12 \pi
 \sin(\frac{\pi}{8})D^{5/4} \ga^{3/4} \left| \Om \right|^{3/4}} .
 \nonumber
  \eea
  The $\Om$-integral gives
  \ben
  \dl \si_T = -
  \left(
  \frac{\mathcal{C}_7 e^2 \al p_F^{1/2}}{4 \pi^2 \sin (\frac{\pi}{8}) D^{1/4} \ga^{3/4}}
  \right)
  T^{1/4},
  \een
  where
  \[
\mathcal{C}_7 \equiv - \int_0^{\infty} dt \frac{\ptl}{\ptl t}
\left( \frac{2t}{e^t -1} \right) \frac{1}{t^{3/4}}
\approx 4.42.
\]
We note that the result given by Eq.~(\ref{eq:t-1over4}) is different from
the result $\dl \sigma_T \propto T^{1/3}$ obtained in Refs.~\onlinecite{prl85}
and~\onlinecite{belitz2}. This is because in the latter the spin fluctuation propagator
is dressed by the non-analytic term proportional to
$\left| q \right|^{d-2} = \left| q \right|$ (which is sub-leading in our model).
It is easy to see from a simple power counting argument that the exponent 1/3 is
obtained by replacing the analytic $(q/p_F)^2$ term by a $\left| q \right|/p_F$ term.

  \emph{Regime V}. (i) For $\dl \ll \ga$ we have
  $U_d^A \approx [\dl + i \ga \Om/(Dq^2)]^{-1}$,
  $B_d^{(3d)} \approx 4/[3 (Dq^2)^2]$, and
  $q_{D3} \sim p_F [\ga \Om/(\eprm \dl)]^{1/2}$ is
  the typical momentum scale. The momentum integral gives
  \bea
& {\rm Im}& \int \frac{d^3 q}{(2 \pi)^3}
 \left[ \dl + \frac{i \ga \Om}{Dq^2} \right]^{-1}
 \frac{4}{3 (Dq^2)^2}
 \nonumber \\
 &=&
 - \frac{{\rm Sgn}(\Om)}{3 \sqrt{2} \pi D^{3/2} (\ga \dl \left| \Om \right|)^{1/2}}.
 \nonumber
 \eea
 (ii) For $\dl \gg \ga$, $U^A \approx 1/\dl$, and the typical momentum scale is
 $q_{D2} \sim p_F(\Om/\eprm)^{1/2}$. Keeping the $\Om$-dependence of
 $B_d^{(3d)}$ we get
 \[
 \frac{1}{\dl}  {\rm Im} \int \frac{d^3 q}{(2 \pi)^3} B_d^{(3d)}
 = - \frac{{\rm Sgn}(\Om)}{8 \sqrt{2} \pi D^{3/2}  \dl \left| \Om \right|^{1/2}}.
 \]
 After the frequency integral we get
 \ben
 \dl \si_T = - \left(
 \frac{\mathcal{C}_8 \mathcal{C}_9 e^2 \al}{\sqrt{2} \pi^2 (\ga \dl D)^{1/2}}
 \right) T^{1/2},
 \een
 where
 \[
 \mathcal{C}_8 \equiv - \int_0^{\infty} dt \frac{\ptl}{\ptl t} \left( \frac{2t}{e^t -1} \right)
 \frac{1}{t^{1/2}} \approx 2.59,
 \]
 and $\mathcal{C}_9 = 1$ for $\dl \ll \ga$, and $\mathcal{C}_9 = (3/8)(\ga/\dl)^{1/2}$
 for $\dl \gg \ga$.

\section{Conclusion}
\label{sec:conclusion}

To conclude, we have calculated the temperature dependence of the conductivity due to
interaction correction for a disordered itinerant electron system close to a ferromagnetic
quantum critical point in dimensions two and three. With an appropriate choice of
parameters $\ga \gg \al$, where $\ga$ is the parameter associated with the Landau damping
of the spin fluctuations and $\al$ is the dimensionless coupling between the conduction
electrons and the spin fluctuations, we are able to perform controlled calculations over the
entire $T$-$\delta$ plane, where $\delta$ is the mass of the spin fluctuations. Near the
quantum critical point, the crossover between diffusive and ballistic regimes of transport
occurs at a temperature $T^{\ast} = 1/[\tau \ga (E_F \tau)^2]$ which is few
orders of magnitude smaller than the crossover temperature $1/\tau$
which is expected in the
case of a good metal (sufficiently far away from any phase instability).
The ballistic-diffusive crossover is determined by the temperature at which the
typical length travelled by an electron during an electron-electron scattering event
becomes comparable with the mean free path $v_F \tau$. Using uncertainty principle,
this length can be estimated as $1/q$, where ${\bf q}$ is the momentum transferred
by the spin fluctuation to the electron. In the quantum critical regime typical $q$
scales as $p_F (\ga \Om/E_F)^{1/3}$ which gives the crossover temperature
$T^{\ast}$ (after scaling $\Om \sim T$). Away from the quantum critical regime,
typical $q$ scales as $\Om/v_F$, which gives the usual crossover scale
$1/\tau$ for a good metal.
In the ballistic regime near the quantum critical point (regime I in Figs.~(4) and (5)),
we obtained $\dl \si_T \propto - T^{(d-1)/3}$, which can be understood from the
following scaling argument. In the ballistic limit
the correction to the transport scattering rate due to
electron-electron interaction can be estimated as $\Delta [1/\tau] \sim (1/\tau)
{\rm Im} \Sigma (\om) \Delta t$, where $\Delta t$ is the average time scale of the
interaction. This can be understood as the renormalization of the quasiparticle
scattering rate ${\rm Im} \Sigma$ by the factor $(\Delta t)/\tau$ in order to obtain
a transport rate. Using the uncertainty principle, we estimate $\Delta t \sim 1/(v_F q)$,
where $q \sim \om^{1/3}$ in the quantum critical regime. Now, since
${\rm Im} \Sigma^{(2d)} \propto \om^{2/3}$ and
${\rm Im} \Sigma^{(3d)} \propto \om$, we get the $T^{(d-1)/3}$ temperature
dependence of the conductivity.
In the diffusive regime near the quantum critical point
(regime IV in Figs.~(4) and (5)) we found
$\dl \si_T \propto \ln^2 T$ in $d=2$ and $\dl \si_T \propto - T^{1/4}$ in $d=3$.
Moving out of the quantum critical regime we recovered the usual results for Fermi
liquids (regimes II and V).

\begin{acknowledgments}

This work was supported by the U. S. Dept. of Energy, Office of Science,
under Contract No. DE-AC02-06CH11357.
The author is very thankful to C. P\'{e}pin, D. L. Maslov, B. N. Narozhny,
I. S. Beloborodov, J. Rech and A. Melikyan for insightful discussions.

\end{acknowledgments}

\appendix

\section{}
\label{appen:a}
In this appendix we give the details of the calculation of the fermionic
component $B^{(2d,3d)}(\bq, \Om)$ in Eq.~(\ref{eq:cond1}) in the ballistic and
the diffusive limits.

\subsection{Ballistic Regime}
In this regime the leading contribution to the conductivity bubble is due
to the diagrams shown in Figs.~(\ref{fig:cond1})
and (\ref{fig:cond2}) (with the solid lines representing disorder averaged electron
Green's functions), i.e.,
those without any explicit impurity scattering, and those with one explicit
impurity scattering respectively.

\subsubsection{$d=2$}

We begin by calculating the diagrams in Fig.~(\ref{fig:cond1}).
Note that
diagrams (c), (d) and (e) have numerical pre-factors -1, -1, and 2
respectively (see their corresponding expressions in
Eq.~(\ref{eq:cond0})). As an example we calculate the diagram in
Fig.~(\ref{fig:cond1}(a)). This is given by
\bea
[B_1^{(2d)}]_{ij}
&=&
\frac{1}{\vol}
\sum_{\bk} G^R(\bk, \om) G^A(\bk,\om) G^R(\bk -\bq, \om - \Om)
\nonumber \\
&\times& G^A(\bk, \om) (v_{\bk})_i (v_{\bk})_j,
\eea
where
$G^{R,A}$ refer to disorder averaged electron Green's functions
in Eq.~(\ref{eq:G-fn}), and $\vol$ is the volume of the system. The
momentum sum in the above Eq. is dominated by the contribution near
the Fermi surface where the spectrum can be linearized, and
$(v_{\bk})_i (v_{\bk})_j = (v_F^2/d) \dl_{ij}$ (assuming an isotropic system).
Replacing
\[
\frac{1}{\vol}\sum_{\bk} \rightarrow \nu_0 \int \frac{d \tht}{2 \pi} \int \ep_k,
\]
where $\nu_0 = m/\pi$, the energy integral is given by
\bea
I_1
&=&
\int_{-\infty}^{\infty} d \ep_k \frac{1}{(\om - \ep_k + \frac{i}{2 \tau})}
\frac{1}{(\om - \ep_k - \frac{i}{2 \tau})^2}
\nonumber \\
&\times&
\frac{1}{(\om - \Om - \ep_k + v_F q \cos \tht + \frac{i}{2 \tau})},
\nonumber
\eea
which can be solved by contour integration. The diagonal component of the
tensor $[B_1^{(2d)}]_{ii} \equiv B_1^{(2d)}$ is given by
\bea
B_1^{(2d)}
&=&
\frac{2 \pi \tau \nu_0 v_F^2}{d} \int_0^{2 \pi}
\frac{d \tht}{2 \pi}
\nonumber \\
& \times &
\left[
\frac{\tau}{(\tOm - i v_F q \cos \tht )} + \frac{1}
{(\tOm - i v_F q \cos \tht )^2} \right]
\nonumber \\
&=&
\prf
\left[
\frac{\tau}{S} + \frac{\tOm}{S^3} \right],
\nonumber
\eea
where $\prf = (2 \pi \tau \nu_0 v_F^2)/d$, and
\bea
\tOm &=& i \Om + \frac{1}{\tau},
\nonumber \\
S &=& \left[ \tOm^2 + (v_F q)^2 \right]^{1/2}.
\eea
The remaining diagrams (b) --- (e) in Fig.~(\ref{fig:cond1})
can be evaluated similarly, giving respectively
\bea
B_2^{(2d)}
&=&
 \prf \left[ - \frac{\tau}{S} \right],
\\
B_3^{(2d)}
&=&
 \prf \left[ \frac{(2 \tOm^2 - (v_F q)^2}{2 \tau S^5} \right],
\\
B_4^{(2d)}
&=&
 \prf \left[ \frac{(v_F q)^2 - 2 \tOm^2}{\tau S^5} \right],
\\
B_5^{(2d)}
&=&
 \prf \left[- \frac{2 \tOm}{S^3} \right].
\eea
The leading behaviour in the ballistic limit is given by the
first non-vanishing term of the expansion of the fermionic part
in the parameter $1/\tau$. We note that $B_1^{(2d)}$ and
$B_2^{(2d)}$ have terms of $\ord (\tau^2)$ which cancel, which
implies that the leading behaviour is due to terms of $\ord (\tau)$.
As a result contributions from $B_3^{(2d)}$ and $B_4^{(2d)}$ are
sub-leading, and can be ignored. The remaining $\ord(\tau)$ contributions
are generated by introducing one explicit impurity
line (represented by dashed line), which gives a factor of
$1/(2\pi \nu_0 \tau)$, in diagrams (a)
and (b) in Fig.~(\ref{fig:cond1}). These second set of contributions are
shown in Fig.~(\ref{fig:cond2}) (note that the diagram in Fig.~(\ref{fig:cond2}(a))
has a factor of $2$ due to symmetry). The contribution from these diagrams
are respectively given by
\bea
B_6^{(2d)} &=& \prf \left[
\frac{2}{S^2} + \frac{2 \tOm}{\tau S^4} \right],
\\
B_7^{(2d)} &=& B_8^{(2d)} = \prf \left[
\frac{1}{(v_F q)^2} \left( 1 - \frac{\tOm}{S} \right)^2 \right],
\\
B_9^{(2d)} &=&  \prf \left[
- \frac{\tOm}{S^3} \right].
\eea
Adding all the terms inside the square brackets in $B_1^{(2d)}, \cdots, B_9^{(2d)}$,
and
considering only the leading term in $1/\tau$ we get Eqs.~(\ref{eq:Bb2})
and (\ref{eq:S0}).

\subsubsection{$d=3$}

In $d=3$ the logic of the evaluation of the fermionic part $B_b^{(3d)}$ is the same
as in the two-dimensional case. The only difference is in the evaluation of  angular
integrals during Fermi surface averages, since in $d=3$
\[
\frac{1}{\vol}
\sum_{\bk} \rightarrow \nu_0 \int_0^{\pi} \frac{d \tht \sin \tht}{2} \int d \ep_k,
\]
where $\nu_0 = (p_F m)/\pi^2$.
For example, the diagram in Fig.~(\ref{fig:cond1}(a)) is now given by
\bea
B_1^{(3d)} &=& \prf \int_0^{\pi} \frac{d \tht}{2} \sin \tht \left[
\frac{\tau}{\tOm - i v_F q \cos \tht} \right.
\nonumber \\
&+&
\left.
 \frac{1}{(\tOm - i v_F q \cos \tht)^2}
\right]
\nonumber \\
&=&
\prf \left[ \frac{\tau}{\tS} + \frac{1}{S^2} \right], \eea where
\ben \frac{1}{\tS} = \frac{1}{2i v_F q} \ln \left[ \frac{\tOm + i
v_F q}{\tOm - i v_F q} \right].
\een
The remaining diagrams are given by
\bea
B_2^{(3d)} &=&  \prf \left[
- \frac{\tau}{\tS} \right],
\\
B_5^{(3d)} &=&  \prf \left[
- \frac{2}{S^2} \right],
\\
B_6^{(3d)} &=& \prf \left[
\frac{2}{\tS^2} + \frac{2}{\tau \tS S^2} \right],
\\
B_7^{(3d)} &=& B_8^{(3d)} =
\prf \left[
\frac{1}{(v_F q)^2} \left( 1 - \frac{\tOm}{\tS} \right)^2 \right],
\\
B_9^{(3d)} &=&  \prf \left[
- \frac{1}{S^2} \right].
\eea
We do not give results for $B_3^{(3d)}$ and $B_4^{(3d)}$ since they are
sub-leading. Adding all the leading $1/\tau$ terms inside the
square brackets in the expressions for $B_1^{(3d)}, \cdots, B_9^{(3d)}$,
we get Eqs.~(\ref{eq:Bb3}) and (\ref{eq:S1}).

\subsection{Diffusive Regime}

In this regime multiple impurity scattering is important. Consequently, the
interaction vertex is dressed by a ladder of impurity
lines (Fig.~\ref{fig:cond3}(c)). Similarly, a single impurity line is
replaced by the same ladder (Fig.~\ref{fig:cond3}(d)). The leading contributions
to the conductivity are due to the diagrams shown in Figs.~(\ref{fig:cond3}(a)) and
(\ref{fig:cond3}(b)). Note that diagram (\ref{fig:cond3}(b)) has a symmetry factor of
$2$.

\subsubsection{$d=2$}
In this regime $1/\tau \gg (i \Om, v_F q)$, so that we get
$S \approx (1/\tau + i \Om + D q^2)$, where $D = v_F^2 \tau/2$ is the diffusion
constant.
The dressed interaction vertex is given by
\bea
\Gamma^{(2d)}(\bq, \Om) &=&
 1 + \Gamma^{(2d)}(\bq, \Om) \frac{1}{2 \pi \nu_0 \tau}
\nonumber \\
& \times &
\frac{1}{\vol}\sum_{\bk} G^A(\bk, \om) G^R(\bk - \bq, \om - \Om),
\nonumber
\eea
evaluating which gives
\ben
\Gamma^{(2d)}(\bq, \Om) = \frac{S}{S - \frac{1}{\tau}}
\approx \frac{1}{\tau} \frac{1}{(i \Om + Dq^2)}.
\een
Similarly, the ladder of impurity lines is given by
\ben
\Lambda^{(2d)}(\bq, \Om) = \frac{\Gamma^{(2d)}(\bq, \Om)}{2 \pi \nu_0 \tau}
\approx
\frac{1}{2 \pi \nu_0 \tau^2} \frac{1}{(i \Om + Dq^2)}.
\een
The diagram in Fig.~(\ref{fig:cond3}(a)) is given by
\[
[B_{10}^{(2d)}]_{ij} = - \frac{1}{2 \pi \nu_0 \tau}
\left(\Gamma^{(2d)} \right)^3 C_i^{(2d)} C_j^{(2d)},
\]
where
\bea
C_i^{(2d)}(\bq, \Om)
&=&
\frac{1}{\vol}\sum_{\bk} G^A(\bk, \om)^2 G^R(\bk - \bq, \om - \Om) (v_{\bk})_i
\nonumber \\
&=&
- (2 \pi \nu_0 v_F) \frac{v_F q}{S^3} \hat{q}_i,
\nonumber
\eea
where $\hat{q}$ is a unit vector. Replacing $\hat{q}_i \hat{q}_j$ by $1/d$
for diagonal components of the conductivity tensor, and using the approximate
forms for $S$ and $\Gamma^{(2d)}$ we get,
\ben
B_{10}^{(2d)} = \prf \left[
- \frac{(v_F q)^2 \tau}{(i \Om + Dq^2)^3} \right].
\een
Similarly, the diagram in Fig.~(\ref{fig:cond3}(b)) is given by
\ben
B_{11}^{(2d)} = \prf \left[
 \frac{2 (v_F q)^2 \tau}{(i \Om + Dq^2)^3} \right].
\een
Adding the terms in the square brackets in $B_{10}^{(2d)}$ and $B_{11}^{(2d)}$
we get $B_d^{(2d)}$ in Eq.~(\ref{eq:Bd}).

\begin{figure}[tbp]
\begin{center}
\includegraphics[width=5cm]{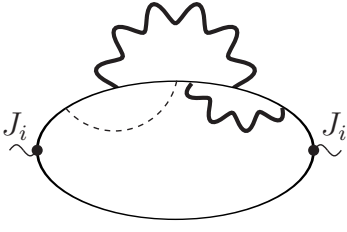}
\end{center}
\caption{An example of a diagram for the conductivity
in the ballistic limit with vertex correction to
the spin-fermion coupling.}
\label{fig:vertex}
\end{figure}

\subsubsection{$d=3$}

Expanding in the small parameters $(i \Om \tau, v_F q \tau)$ we get
$\tS \approx (1/\tau + i \Om + Dq^2)$, with the diffusion constant
in three dimensions given by $D = (v_F^2 \tau)/3$. It is easy to check
that the approximate expressions for $\Gamma^{(3d)}$ and $\Lambda^{(3d)}$
are the same as those in two dimensions (with the appropriate re-definition
of the diffusion constant). Next, we get
\[
C_i^{(3d)}(\bq, \Om)= - (2 \pi \nu_0 v_F) \frac{1}{v_F q} \left(
\frac{1}{\tS} - \frac{\tOm}{S^2} \right) \hat{q}_i.
\]
Using the approximate forms of $\tS$, $S$ and $\Gamma^{(3d)}$ we get
\bea
B_{10}^{(3d)}
&=&
\prf \left[
- \frac{4}{9} \frac{(v_F q)^2 \tau}{(i \Om + Dq^2)^3} \right],
\\
B_{11}^{(3d)}
&=&
\prf \left[
 \frac{8}{9} \frac{(v_F q)^2 \tau}{(i \Om + Dq^2)^3} \right].
\eea
Adding the terms in the square brackets in the above two Eqs.
we get $B_d^{(3d)}$ in Eq.~(\ref{eq:Bd}).

\section{}
\label{appen:b}

In this appendix we discuss the effect of adding a vertex correction to the spin-fermion
coupling in the conductivity calculation for the ballistic limit of the
quantum critical regime. Using Eq.~(\ref{eq:inequality}) we show that, even though the spin fluctuation is
massless, a vertex correction gives rise to sub-leading contribution to the
temperature dependence of conductivity. We demonstrate this explicitly for the
diagram shown in Fig.~(\ref{fig:vertex}), which is a typical example. The behaviour of other vertex
correction diagrams are expected to be similar.

\subsection{$d=2$}

In order to facilitate the discussion we will first evaluate the diagram shown in
Fig.~(\ref{fig:cond2}(a)) in Matsubara frequency, and then compare it with the
evaluation of the corresponding vertex diagram (Fig.~(\ref{fig:vertex})). Writing the
contribution of the former
to the current-current correlator as $[\Pi_6^{(2d)}]_{ij}$ we get,
\bea
\lefteqn{
[\Pi_{6}^{(2d)}(i \nu_n)]_{ij}
=
\left( - \frac{4}{2 \pi \nu_0 \ta} \right)
\left( \frac{3 \al}{\nu_0} \right)\frac{1}{\be^2} \sum_{\om_n, \Om_n}
\frac{1}{\vol^3} \sum_{\bk, \bq, \bp}
}
\nonumber \\
&& \times
 \left(v_{\bk} \right)_i \left(v_{\bk} \right)_j
U(\bq, i \Om_n) \mG(\bk, i \om_n)^2 \mG(\bk, i \om_n + i \nu_n)
\nonumber \\
&& \times
\mG(\bk + \bq, i\om_n + i \Om_n)
\mG(\bp, i \om_n)
\mG(\bp + \bq, i \om_n + i \Om_n),
\nonumber
\eea
where $\om_n$ is fermionic Matsubara frequency, $\Om_n$ and $\nu_n$
are bosonic Matsubara frequencies, and
$\mG$ is the disorder averaged electron Green's function in Matsubara
frequency.  The factor $4$ in front is due to the symmetry of the diagram.
The $\ep_{\bp}$ integral in the above expression is non-zero only if $\om_n$
and $(\om_n + \Om_n)$ have opposite signs. Next, for the $\ep_{\bk}$ integral
the dominant contribution occurs when $\om_n$ and $(\om_n + \nu_n)$
have opposite signs. For $\nu_n > 0$, the leading term in $1/(v_F q \tau)$
is given by,
\bea
\lefteqn{
[\Pi_{6}^{(2d)}(i \nu_n)]_{ii}
=
-  \frac{6 \al v_F^2}{\tau \left( \nu_n + \frac{1}{\ta} \right)^2}
 \frac{1}{\vol} \sum_{\bq}
  \left[
  \frac{1}{\be}
 \sum_{\Om_n > \nu_n} \nu_n
 \right.
 }
 \nonumber \\
 &&
 \left.
 +
   \frac{1}{\be}
 \sum_{0 \leq \Om_n \leq \nu_n} \Om_n
 \right]
 \frac{U(\bq, i \Om_n)}{\left[
 \Om_n^2 + (v_F q)^2
 \right]}.
 \eea
\begin{figure}[tbp]
\begin{center}
\includegraphics[width=8cm]{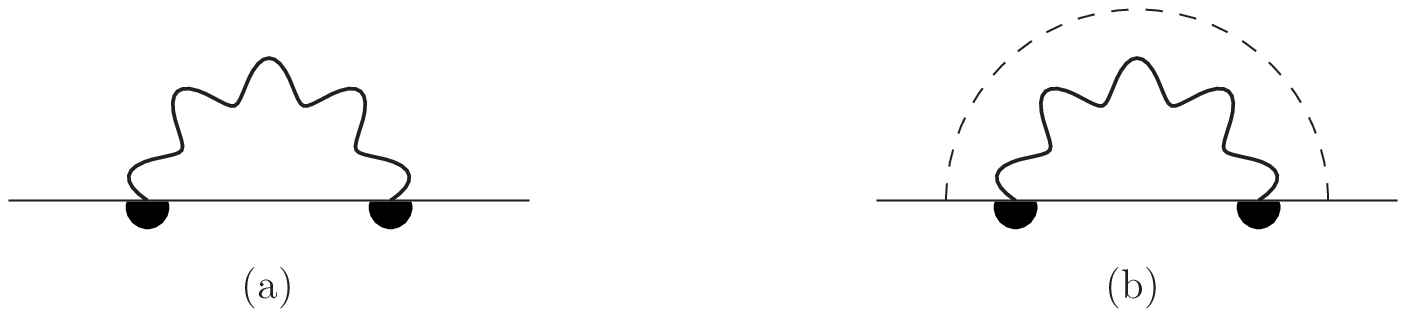}
\end{center}
\caption{Leading diagrams for the self-energy correction in the diffusive
limit of the quantum critical regime.}
\label{fig:diffusiveselfenergy}
\end{figure}
After performing the $\Om_n$ sum, and taking the limit of static conductivity,
we get Eq.~(\ref{eq:cond1})
with $B^{(2d)}(\bq, \Om) = 2/[(v_Fq)^2 - \Om^2 + i \eta {\rm Sgn}(\Om)]$. This is
the contribution to conductivity from the diagram in Fig.~(\ref{fig:cond2}(a))
[see expression for $B_6^{(2d)}$ in Appendix~\ref{appen:a}].
The evaluation of the vertex diagram (Fig.~(\ref{fig:vertex})) is analogous. Writing its contribution to
the current-current correlator as $[\Pi_{6v}^{(2d)}]_{ij}$ we get,
\bea
\lefteqn{
[\Pi_{6v}^{(2d)}(i \nu_n)]_{ij}
=
\left( - \frac{4}{2 \pi \nu_0 \ta} \right)
\left( \frac{3 \al}{\nu_0} \right)^2 \frac{1}{\be^2} \sum_{\om_n, \Om_n}
\frac{1}{\vol^3} \sum_{\bk, \bq, \bp}
}
\nonumber \\
&& \times
 \left(v_{\bk} \right)_i \left(v_{\bk} \right)_j
U(\bq, i \Om_n) \mG(\bk, i \om_n)^2 \mG(\bk, i \om_n + i \nu_n)
\nonumber \\
&& \times
\mG(\bk + \bq, i\om_n + i \Om_n)
\mG(\bp, i \om_n)
\mG(\bp + \bq, i \om_n + i \Om_n),
\nonumber \\
&& \times
\frac{1}{\be \vol} \sum_{r_n, \bq_1}
U(\bq_1, i r_n) \mG(\bk + \bq_1, i \om_n + i r_n)
\nonumber \\
&& \times
\mG(\bk + \bq + \bq_1, i \om_n + i \Om_n + i r_n).
\nonumber
\eea
As in the case of $\Pi_{6}^{(2d)}$, the leading contribution comes from
$\om_n < 0$, $(\om_n + \Om_n) > 0$, and $(\om_n + \nu_n) > 0$.
Taking into account only the leading dependence in $1/\tau$
we get (for $\nu_n >0$),
\bea
\lefteqn{
[\Pi_{6v}^{(2d)}(i \nu_n)]_{ii}
=
-  \frac{18 \al^2 v_F^2 \ta}{\nu_0}
 \frac{1}{\vol} \sum_{\bq}
  \left[
  \frac{1}{\be}
 \sum_{\Om_n > \nu_n} \nu_n
 \right.
 }
 \nonumber \\
 &&
 \left.
 +
   \frac{1}{\be}
 \sum_{0 \leq \Om_n \leq \nu_n} \Om_n
 \right]
 \frac{U(\bq, i \Om_n)}{\sqrt{
 (\Om_n )^2 + (v_F q)^2}}
 \nonumber \\
 && \times
 \int_0^{2 \pi} \frac{d \theta}{2 \pi}
 \frac{M(\Om_n, q \cos \theta)}
 {(\Om_n + i v_F q \cos \theta)^2}.
 \nonumber
 \eea
 Here
 \bea
 M(\Om_n, q \cos \theta)
 &=&
 \frac{1}{\be} \sum_{r_n} \int_0^{\infty} \frac{d q_1 q_1}{2 \pi}
 \int_0^{2 \pi} \frac{d \theta_1}{2 \pi}
 \nonumber \\
 & \times &
 \frac{U(\bq_1, i r_n)}{\Om_n + r_n + i v_F q_1 \cos \theta_1 + i v_F q \cos
 \theta}
 \nonumber \\
 & \approx &
 \frac{p_F^2}{\pi^2} \left[ \frac{0.3 \Om_n^{2/3}}{2 \ga^{1/3} (p_F v_F)^{2/3}}
 - \frac{i q \cos \theta}{6 \ga p_F} \right],
 \nonumber
 \eea
at zero temperature.~\cite{rech0} Using the above expression and performing the angle
integration we finally get,
\bea
\lefteqn{
[\Pi_{6v}^{(2d)}(i \nu_n)]_{ii}
=
 \frac{18 \al^2 v_F^2 p_F^2 \ta}{ \pi^2 \nu_0}
 \frac{1}{\vol} \sum_{\bq}
  \left[
  \frac{1}{\be}
 \sum_{\Om_n > \nu_n} \nu_n
 \right.
 }
 \nonumber \\
 &&
 \left.
 +
   \frac{1}{\be}
 \sum_{0 \leq \Om_n \leq \nu_n} \Om_n
 \right]
 \frac{U(\bq, i \Om_n)}{\left[
 (\Om_n )^2 + (v_F q)^2 \right]^2}
 \nonumber \\
 && \times
 \left\{
 \frac{(v_F q)^2}{6 \ga p_F v_F}
 - \frac{3 \Om_n^{5/3}}{20 \ga^{1/3} (p_F v_F)^{2/3}}
 \right\}.
 \eea
Comparing the expressions for $\Pi_6^{(2d)}$ and $\Pi_{6v}^{(2d)}$ we get
that in regime I, where $(q/p_F) \sim (\ga \Om/E_F)^{1/3}$, the vertex correction
also yields $\dl \si \propto - T^{1/3}$, but with a pre-factor which is parametrically
small in $(\al/\ga)$. In the high temperature regime of the theory (regime III),
where $q \sim \Om/v_F$, the vertex correction is small by
$(\al/\ga^{1/2})(T_1/T)^{1/3}$.

\subsection{$d=3$}

\begin{figure}[tbp]
\begin{center}
\includegraphics[width=3cm]{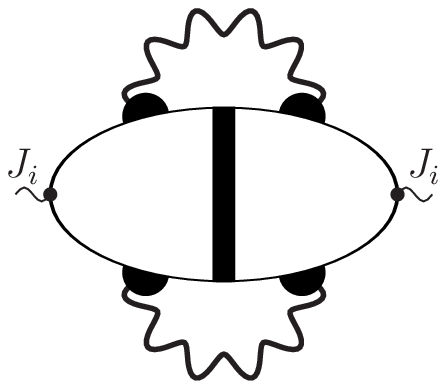}
\end{center}
\caption{An example of a diagram for the conductivity
in the diffusive limit at second order in
the spin-fermion coupling.}
\label{fig:diffusivevertex}
\end{figure}

In three dimensions the estimation of the vertex diagram is entirely analogous to the
$d=2$ case. The contribution to the current-current correlator from the diagram shown in
Fig.~(\ref{fig:cond2}(a)) is given by
\bea
\lefteqn{
[\Pi_{6}^{(3d)}(i \nu_n)]_{ii}
=
-  \frac{4 \al v_F^2}{\tau \left( \nu_n + \frac{1}{\ta} \right)^2}
 \frac{1}{\vol} \sum_{\bq}
  \left[
  \frac{1}{\be}
 \sum_{\Om_n > \nu_n} \nu_n
 \right.
 }
 \nonumber \\
 &&
 \left.
 +
   \frac{1}{\be}
 \sum_{0 \leq \Om_n \leq \nu_n} \! \! \! \! \Om_n
 \right]
 \!
 \frac{U(\bq, i \Om_n)}{(v_F q)^2}
 \!
 \left[
 \tan^{-1} \left( \frac{v_F q}{\Om_n} \right)
 \right]^2.
\eea
On the other hand, the corresponding contribution from the vertex diagram of
Fig.~(\ref{fig:vertex}) is given by
\begin{widetext}
\bea
[\Pi_{6v}^{(3d)}(i \nu_n)]_{ii}
&=&
 \frac{12 \al^2 v_F p_F^2 \ta}{\nu_0}
 \frac{1}{\vol} \sum_{\bq}
  \left[
  \frac{1}{\be}
 \sum_{\Om_n > \nu_n} \nu_n
 +
 \frac{1}{\be}
 \sum_{0 \leq \Om_n \leq \nu_n} \Om_n
 \right]
 \frac{U(\bq, i \Om_n)}{v_F q}
 \tan^{-1} \left( \frac{v_F q}{\Om_n} \right)
 \nonumber \\
 & \times &
 \left\{
 \frac{1}{2 \sqrt{3} \pi^2 \ga^{1/3} v_F q}
 \tan^{-1} \left(\frac{v_F q}{\Om_n} \right)
 +
 \frac{\Om_n}{(v_F q)^2 + \Om_n^2} \left( \frac{1}{12 \pi^2} \right)
 \ln \left( \frac{p_F v_F}{\Om_n} \right)
 \right\}.
 \eea
\end{widetext}
Comparing the expressions for $\Pi_6^{(3d)}$ and $\Pi_{6v}^{(3d)}$ we get
that in regime I, where $(q/p_F) \sim (\ga \Om/E_F)^{1/3}$, the vertex correction
is parametrically
small in $(\al/\ga^{1/3})$. In the high temperature regime of the theory (regime III),
where $q \sim \Om/v_F$, the vertex correction is small in
$\al \ln(E_F/T)$.

\section{}
\label{appen:c}

In this appendix we calculate the leading self-energy correction of the electron propagator
in the diffusive limit of the quantum critical regime. This is  given by the diagrams shown
in Fig.~(\ref{fig:diffusiveselfenergy}). It is to be noted that there are four other contributions
which are important for the self-energy away from the quantum critical
regime where FL results are recovered (i.e., in Regime V of Figs.~(\ref{fig:scales1}) and
(\ref{fig:scales2})),~\cite{adamov} but which are sub-leading in the quantum critical regime.
 Using Eq.~(\ref{eq:inequality}) we show
that the self-energy correction is much smaller than the elastic scattering rate $1/\tau$, and
therefore such a correction can be omitted in a perturbative calculation.
Furthermore, we argue that the terms which are second order in the spin-fermion coupling
are sub-leading in the diffusive quantum critical regime due to Eq.~(\ref{eq:inequality}). We
demonstrate this explicitly for the diagram shown in Fig.~(\ref{fig:diffusivevertex}). The
behaviour of the other second order terms are expected to be similar or smaller.

\subsection{$d=2$}

First we evaluate the self-energies given by Fig.~(\ref{fig:diffusiveselfenergy}).
Denoting the diagram (a) as $\Sigma_{1}^{(2d)}(i \om_n, \bk)$, we have
\bea
\Sigma_{1}^{(2d)}(i \om_n, \bk)
&=&
\left( \frac{3 \al}{\nu_0} \right) \frac{1}{\be} \sum_{\Om_n}^{\prime}
\frac{1}{\vol} \sum_{\bq} U_d(\bq, i\Om_n)
\nonumber \\
&\times& \Gamma^2 (\bq, i \Om_n) \mG(\bk + \bq, i \om_n + i \Om_n),
\nonumber
\eea
where the prime in the frequency summation indicates the condition that
$(\om_n + \Om_n) \om_n < 0$, and where
$\Gamma (\bq, i \Om_n) = 1/[\tau ( D q^2 + \left| \Om_n \right|)]$ in the
diffusive limit. We evaluate the above expression at the pole of the
electron Green's function,
i.e., at $\bk$ such that $\mG^{-1} (\bk, i \om_n) = 0$.~\cite{adamov}
This gives
\bea
\Sigma_{1}^{(2d)}(i \om_n)
&=&
i \; {\rm Sgn} (\om_n)
\left( \frac{3 \al}{\nu_0} \right) \frac{1}{\be} \sum_{\Om_n}^{\prime}
\frac{1}{\vol} \sum_{\bq}
\nonumber \\
&\times&
\frac{U_d(\bq, i\Om_n) \Gamma^2 (\bq, i \Om_n)}
{\sqrt{(\left| \Om_n \right| + 1/\tau)^2 + (v_F q)^2}}.
\nonumber
\eea
In the above the leading term in an expansion in $\tau$ is divergent, but this divergence is
canceled by the diagram (b) in Fig.~(\ref{fig:diffusiveselfenergy}). The latter contribution,
which is momentum independent is given by,
\bea
\Sigma_{2}^{(2d)}(i \om_n)
&=&
\left( \frac{3 \al}{\nu_0} \right) \frac{1}{\be} \sum_{\Om_n}^{\prime}
\frac{1}{\vol^2} \sum_{\bq, \bp} U_d(\bq, i\Om_n) \Gamma^2 (\bq, i \Om_n)
\nonumber \\
&\times& \frac{1}{2 \pi \nu_0 \tau}
\mG^2 (\bp, i \Om_n) \mG(\bp + \bq, i \om_n + i \Om_n)
\nonumber \\
&=&
- i \; {\rm Sgn} (\om_n)
\left( \frac{3 \al}{\nu_0} \right) \frac{1}{\be} \sum_{\Om_n}^{\prime}
\frac{1}{\vol} \sum_{\bq} U_d(\bq, i\Om_n)
\nonumber \\
& \times &
\Gamma^2 (\bq, i \Om_n) \frac{1/\tau (\left| \Om_n \right| + 1/\tau )}
{\left[ (\left| \Om_n \right| + 1/\tau)^2 + (v_F q)^2 \right]^{3/2} }.
\nonumber
\eea
In the diffusive limit of the quantum critical regime we have
$1/\tau \gg D q^2 \gg \left| \Om_n \right|$, using which we get
\ben
\label{eq:diffusiveselfenergy}
\begin{split}
\Sigma_{d}^{(2d)}(i \om_n)
& =
\Sigma_{1}^{(2d)}(i \om_n) + \Sigma_{2}^{(2d)}(i \om_n)
\\
&=
- i \; {\rm Sgn} (\om_n) \left( \frac{3 \al p_F}{4 \pi \nu_0 D^{1/2} \ga^{1/2}} \right)
\left| \om_n \right|^{1/2},
\end{split}
\een
to the lowest order in $\tau$. In the above we ignored a constant part coming from the
ultraviolet cut-off. Scaling $\left| \om_n \right| \sim T$, and using Eq.~(\ref{eq:inequality})
we conclude that for $T \ll T^{\ast}$, $1/\tau \gg \Sigma_{d}^{(2d)}$. Thus, the self-energy
correction can be ignored for the evaluation of the leading temperature dependence of the
conductivity.

Next, we evaluate the contribution to the current-current correlator from the diagram shown in
Fig.~(\ref{fig:diffusivevertex}).
For this we define
\bea
C_i^{(2d)}
&=&
\frac{1}{\vol} \sum_{\bp} (v_{\bp})_i \mG (\bp, i \om_n) \mG (\bp + \bq, i \om_n + i \Om_n)
\nonumber \\
&\times&
\mG (\bp, i \om_n + i \nu_n) \mG (\bp + \bq_1, i \om_n + i \nu_n + i \Om^{\prime}_n)
\nonumber \\
& \approx &
(4 \pi i \nu_0 v_F^2 \tau^4) (\bq_1 - \bq)_i,
\nonumber
\eea
and $\Lambda(\bq, i \Om_n) = \Gamma (\bq, i\Om_n)/(2 \pi \nu_0 \tau)$.
Then the current-current correlator can be written as
\bea
[\Pi_{v}^{(2d)}(i \nu_n)]_{ij}
&=&
- \frac{1}{\vol^2} \sum_{\bq , \bq_1} \frac{1}{\be^3} \sum_{\om_n, \Om_n, \Om_n^{\prime}}
\left( \frac{3 \al}{\nu_0} \right)^2 C^{(2d)}_i
\nonumber \\
&\times&
C^{(2d)}_j
\Gamma^2(\bq, i \Om_n) U_d (\bq, i \Om_n)
\Gamma^2(\bq_1, i \Om_n)
\nonumber \\
&\times&
U_d (\bq_1, i \Om_n)
\Lambda (\bq - \bq_1, i \Om_n - i \Om_n^{\prime} - i \nu_n).
\nonumber
\eea
In the above the restrictions on the frequency summations are such that, for
$\nu_n > 0 $, we have $(\om_n + \Om_n) > 0$, $(\om_n + \nu_n) > 0$, $\om_n < 0$
and $(\om_n + \nu_n + \Om_n^{\prime}) < 0$. For the purpose of an estimation, we
perform the $\Om_n^{\prime}$-summation without restriction, and we get
\bea
[\Pi_{v}^{(2d)}(i \nu_n)]_{ii}
& \approx &
\left( \frac{9 \al^2 v_F^2 \tau}{8 \pi^2 \ga \nu_0  D} \right)
\ln^2 \left( \frac{T}{\ga D p_F^2} \right) \frac{1}{\vol} \sum_{\bq}
\nonumber \\
& \times &
\left[ \frac{1}{\be} \sum_{\Om_n > \nu_n} \nu_n +
\frac{1}{\be} \sum_{0 \leq \Om_n \leq \nu_n} \Om_n
 \right]
 \nonumber \\
& \times &
 \frac{U_d(\bq, i \Om_n)}{(D q^2)^2}.
\eea
By comparing with a typical contribution to the current-current correlator at
first order in $\al$, we conclude that the contribution from the diagram shown
in Fig.~(\ref{fig:diffusivevertex}) is smaller by a factor of
$\al \ln ^2 [T/(\ga D p_F^2)]/(\ga E_F \ta)$.

\subsection{$d=3$}

In three dimensions the calculations are entirely analogous to the $d=2$ case.
For the self-energy given by Fig.~(\ref{fig:diffusiveselfenergy}) we get
\ben
\Sigma_{d}^{(3d)}(i \om_n)
\approx
- i \; {\rm Sgn} (\om_n) \left( \frac{\al p_F^{3/2}}
{\sqrt{2} \pi^2 \nu_0 D^{3/4} \ga^{1/4}} \right)
\left| \om_n \right|^{3/4},
\een
which is smaller than $1/\tau$ for $T \ll T^{\ast}$, and therefore can be neglected.
Next, we estimate the contribution to the current-current correlator
from the diagram in Fig.~(\ref{fig:diffusivevertex}), and we find that it is smaller
than those that are first order in $\al$ by a factor of
$\al T^{1/4}/(\ga^{3/4} \tau^{5/4} E_F^{3/2})$.


\begin{thebibliography}{99}
\bibitem{altshuler}  B. L. Altshuler, and A. G. Aronov,
\textit{Electron-Electron Interactions in Disordered Systems},
edited by A. L. Efros
and M. Pollak (North-Holland, Amsterdam, 1985), and references therein.

\bibitem{aleiner1} I. L. Aleiner, B. L. Altshuler,
and M. E. Gershenson, Waves Random Media \textbf{9}, 201 (1999).

\bibitem{zna}  G. Zala, B. N. Narozhny, and I. L. Aleiner, Phys. Rev. B
\textbf{64}, 214204 (2001).

\bibitem{stewart} for a review see e.g.,
G. R. Stewart, Rev. Mod. Phys. \textbf{73}, 797 (2001).

\bibitem{ipaul} I. Paul, C. P\'{e}pin, B. N. Narozhny, and D. L. Maslov,
Phys. Rev. Lett. \textbf{95}, 017206 (2005).

\bibitem{kim}  Y. B. Kim, and A. J. Millis, Phys. Rev. B \textbf{67}, 085102
(2003).

\bibitem{prl85} D. Belitz, T. R. Kirkpatrick, R. Narayanan, and T. Vojta,
Phys. Rev. Lett. \textbf{85}, 4602 (2000).

\bibitem{belitz2}  D. Belitz, T. R. Kirkpatrick, M. T. Mercaldo, and S. L.
Sessions, Phys. Rev. B \textbf{63}, 174428 (2001).

\bibitem{rosch1}  A. Rosch, Phys. Rev. Lett. \textbf{82}, 4280 (1999).

\bibitem{huxley} C. Pfleiderer, and A. D. Huxley, Phys. Rev. Lett. \textbf{89},
147005 (2002).

\bibitem{uhlarz} M. Uhlarz, C. Pfleiderer, and S. M. Hayden, Phys.
Rev. Lett. \textbf{93}, 256404 (2004).

\bibitem{grigera}
S. A. Grigera, R. S. Perry, A. J. Schofield, M. Chiao, S. R. Julian, G. G. Lonzarich,
S. I. Ikeda, Y. Maeno, A. J. Millis, and A. P. Mackenzie,
Science \textbf{294}, 329 (2001).

\bibitem{schofield}  A. J. Schofield, A. J. Millis, S. A. Grigera, and G. G.
Lonzarich, in Springer Lecture Notes in Physics, \textbf{603}, 271 (2002).

\bibitem{grigera2}
S. A. Grigera, P. Gegenwart, R. A. Borzi, F. Weickert, A. J. Schofield, R. S. Perry,
T. Tayama, T. Sakakibara, Y. Maeno, A. G. Green, and A. P. Mackenzie,
Science \textbf{306}, 1154 (2004).

\bibitem{sokolov} D. A. Sokolov, M. C. Aronson, W. Gannon, and Z. Fisk,
Phys. Rev. Lett. \textbf{96}, 116404 (2006).

\bibitem{lee}  P. A. Lee, and T. V. Ramakrishnan,
Rev. Mod. Phys. \textbf{57}, 287 (1985).

\bibitem{mirlin} A. D. Mirlin, and P. W\"{o}lfle, Phys. Rev. B
\textbf{55}, 5141 (1997).

\bibitem{khveshchenko} D. V. Khveshchenko, Phys. Rev. Lett.
\textbf{77}, 362 (1996).

\bibitem{galitski} V. M. Galitski, Phys. Rev. B  \textbf{72}, 214201 (2005).

\bibitem{hmm}  J. A. Hertz, Phys. Rev. B \textbf{14}, 1165 (1976);
T. Moriya, \textit{Spin Fluctuations in Itinerant Electron
Magnetism}, (Springer-Verlag, Berlin, New York, 1985);
A. J. Millis, Phys. Rev. B \textbf{48}, 7183 (1993).

\bibitem{belitz1}  D. Belitz, T. R. Kirkpatrick, and T. Vojta, Phys. Rev. B
\textbf{55}, 9452 (1997).

\bibitem{abanov}  Ar. Abanov, and A. V. Chubukov,
Phys. Rev. Lett. \textbf{93}, 255702 (2004).

\bibitem{belitz-rmp} D. Belitz, T. R. Kirkpatrick, and T. Vojta,
Rev. Mod. Phys. \textbf{77}, 579 (2005).

\bibitem{chubukov1}  Ar. Abanov, A. V. Chubukov, and J. Schmalian, Adv.
Phys. \textbf{52}, 119 (2003).

\bibitem{agd}  A. A. Abrikosov, L. P. Gorkov, and I. E. Dzyaloshinski,
\textit{Methods of Quantum Field Theory in Statistical Physics}, (Dover
Publications Inc., New York, 1963).

\bibitem{infrared} A. V. Chubukov, and D. L. Maslov, Phys. Rev. B \textbf{68},
155113 (2003); G. Y. Chitov, and A. J. Millis, Phys. Rev. Lett. \textbf{86},
5337 (2001).

\bibitem{rech}  A. V. Chubukov, C. P\'{e}pin, and J. Rech, Phys. Rev. Lett.
\textbf{92}, 147003 (2004).

\bibitem{disorder-belitz} T. R. Kirkpatrick, and D. Belitz,
Phys. Rev. B \textbf{53}, 14364 (1996).

\bibitem{kamenev} A. Kamenev, and Y. Oreg, Phys. Rev. B {\bf 52}, 7516 (1995).

\bibitem{rech0} J. Rech, C. P\'{e}pin, and A. V. Chubukov,
Phys. Rev. B {\bf 74}, 195126 (2006).

\bibitem{adamov} Y. Adamov, I. V. Gornyi, and A. D. Mirlin,
Phys. Rev. B {\bf 73}, 045426 (2006).



\end{thebibliography}
\end{document}